\documentclass[10pt,aps,prd,twocolumn,showpacs,amsmath,nofootinbib]{revtex4-2}

\usepackage[utf8]{inputenc}
\usepackage[american]{babel}
\usepackage[a4paper,margin=1.8cm]{geometry}

\usepackage{amsmath,amssymb}
\usepackage{booktabs}
\usepackage{graphicx}
\usepackage[labelfont=bf,font+=smaller,justification=centerlast]{caption}
\usepackage[labelfont=normal,font=normal,font=small,labelformat=simple]{subcaption}
\usepackage{xfrac}
\usepackage{multirow}
\usepackage{mathtools}
\usepackage{verbatim}
\usepackage[absolute,overlay]{textpos}

\graphicspath{{./}{./diagrams/}}

\usepackage{hyperref}
\usepackage{cleveref}
\hypersetup{%
colorlinks,%
citecolor=blue,urlcolor=blue,linkcolor=blue,%
hypertexnames=false,%
pdftitle={Six-meson amplitude in QCD-like theories},%
pdfauthor={J. Bijnens, T. Husek, M. Sjo}%
}

\usepackage{scalerel,stackengine}
\newcommand\pig[1]{\scalerel*[5.5pt]{\Big#1}{%
  \ensurestackMath{\addstackgap[1.5pt]{\big#1}}}}

\renewcommand{\cref}[1]{\Cref{#1}}

\Crefname{equation}{Eq.}{Eqs.}
\Crefname{figure}{Fig.}{Figs.}
\Crefname{tabular}{Tab.}{Tabs.}
\Crefname{section}{Sec.}{Secs.}

\newcommand{\rcite}[1]{Ref.~\cite{#1}}
\newcommand{\rrcite}[1]{Refs.~\cite{#1}}

\newcommand{\RRcite}[1]{References~\cite{#1}}

\renewcommand{\thesubsection}{\Roman{section}--\Alph{subsection}}
\makeatletter
\let\p@subsection
\makeatother

\newcommand{\f}{b}
\newcommand{\N}{k}
\newcommand{\R}{\mathrm{\dag}}
\newcommand{\T}{\mathsf{T}}
\newcommand{\lagr}{{\mathcal L}}
\newcommand{\flav}{{\mathcal F}}
\newcommand{\mandel}[1]{\hat s_{#1}}

\newcommand{\tr}[1]{\langle#1\rangle}
\renewcommand{\r}{\mathrm{r}}
\newcommand{\rev}{\text{\texttt{+}\sc{tr}}}
\newcommand{\uu}{\hat U}

\newcommand{\ampl}{{A}}
\newcommand{\sampl}[1]{\ampl_{\{#1\}}}
\newcommand{\dampl}{\tilde\ampl}

\newcommand{\gampl}[1]{{\mathcal\ampl}^{(#1)}}
\newcommand{\dgampl}[1]{\tilde{\mathcal\ampl}^{(#1)}}
\newcommand{\sgampl}[2]{\gampl{#2}_{\{#1\}}}
\newcommand{\dsgampl}[2]{\tilde{\mathcal\ampl}_{\{#1\}}^{(#2)}}
\newcommand{\hampl}[0]{{\hat\ampl}}

\newcommand{\hsgampl}[2]{{\hat{\mathcal\ampl}}_{\{#1\}}^{(#2)}}
\newcommand{\graph}[1]{\mathcal{M}_\text{#1}}
\newcommand{\ograph}[2]{\graph{#1}^{(#2)}}

\newcommand{\pole}{\text{pole}}
\newcommand{\nonpole}{\text{nonpole}}
\newcommand{\LO}{\text{LO}}
\newcommand{\NLO}{\text{NLO}}
\newcommand{\sgn}{\xi}
\newcommand{\sgnp}{\xi^2}
\newcommand{\kk}{\zeta}

\DeclareMathOperator{\id}{id}
\newcommand{\Z}{\mathbb{Z}}
\DeclareMathOperator{\SU}{SU}
\DeclareMathOperator{\UU}{U}
\DeclareMathOperator{\SO}{SO}
\DeclareMathOperator{\OO}{O}
\DeclareMathOperator{\Sp}{Sp}
\DeclareMathOperator{\SOp}{S\overset{\lower.55em\hbox{\scalebox{0.55}{$\mathrm O$}}}{\lower.03em\hbox{\scalebox{0.6}{$\mathrm p$}}}}

\DeclareMathAlphabet{\mathbbb}{U}{bbold}{m}{n}
\newcommand{\1}{\mathbbb{1}}

\newlength{\eqindent}
\newlength{\peqindent}
\newlength{\ppeqindent}
\newlength{\pppeqindent}
\newcommand{\eqbreak}[1]{%
	\notag\\&\hspace{\eqindent}#1\,}
\newcommand{\peqbreak}[1]{%
	\notag\\&\hspace{\peqindent}#1\,}
\newcommand{\ppeqbreak}[1]{%
	\notag\\&\hspace{\ppeqindent}#1\,}
\newcommand{\pppeqbreak}[1]{%
	\notag\\&\hspace{\pppeqindent}#1\,}

\newcommand\scalemath[2]{\scalebox{#1}{\normalsize\mbox{\ensuremath{\displaystyle #2}}}}
\newcommand\smallfrac[2]{\tfrac{\raisebox{1pt}{\scriptsize\mbox{\ensuremath{#1}}}}{\raisebox{-1pt}{\scriptsize\mbox{\ensuremath{#2}}}}}

\usepackage{pgf,tikz}
\usetikzlibrary{intersections, calc}

\tikzset{
    prop/.style={thick,join=round},
    singlet/.style={prop,densely dashed}
}

\allowdisplaybreaks % because we have some very long equations

\begin{document}

\title{Six-meson amplitude in QCD-like theories}
\author{Johan Bijnens}
\email{johan.bijnens@thep.lu.se}
\author{Tom\'{a}\v{s} Husek}
\email{tomas.husek@thep.lu.se}
\author{Mattias Sj\"o}
\email{mattias.sjo@thep.lu.se}
\affiliation{Department of Astronomy and Theoretical Physics, Lund University,\\ Box 43, SE 221-00 Lund, Sweden}
\date{\today}

\begin{abstract}
We calculate the relativistic six-meson scattering amplitude at low energy within the framework of QCD-like theories with $n$ degenerate quark flavors at next-to-leading order in the chiral counting. We discuss the cases of complex, real and pseudo-real representations, i.e.\ with global symmetry and breaking patterns $\SU(n)\times\SU(n)/\SU(n)$ (extending the QCD case), $\SU(2n)/\SO(2n)$, and $\SU(2n)/\Sp(2n)$. In case of the one-particle-irreducible part, we obtain analytical expressions in terms of 10 six-meson subamplitudes based on the flavor and group structures. We extend on our previous results obtained within the framework of the $\OO(N+1)/\OO(N)$ non-linear sigma model, with $N$ being the number of meson flavors. This work allows for studying a number of properties of six-particle amplitudes at one-loop level.
\end{abstract}

\pacs{
12.39.Fe Chiral Lagrangians,
11.30.Rd Chiral symmetries,
14.40.Aq pi, $K$, and eta mesons\\[1mm]
PhysH: Chiral perturbation theory, Effective field theory, Scattering amplitudes, Nonlinear sigma model, Chiral symmetry}

\maketitle

\begin{textblock*}{\paperwidth}(-\leftmargin,1cm)
   \hfill LU TP 22-45
\end{textblock*}

%%%%%%%%%%%%%%%%%%%%%%%%%%%%%%%%%%%%%%%%%%%%%%%%%%%%%%%%%%%%%%%%%%%%%%%%%
\section{Introduction}

Quantum chromodynamics (QCD), the fundamental theory of the strong interaction, becomes non-perturbative at low energy and it is therefore impractical for phenomenology in that regime.
From the large-distance perspective, the fundamental quark and gluon degrees of freedom are effectively replaced by composite colorless states, the lightest of which are the mesons.
These can be approximately interpreted as the Nambu--Goldstone bosons of the associated spontaneous breaking of the chiral symmetry of massless QCD.
With appropriate explicit symmetry breaking added to account for quark masses and non-strong interactions, the resulting effective field theory (EFT) is known as chiral perturbation theory (ChPT)~\mbox{\cite{Weinberg:1978kz,Gasser:1983yg,Gasser:1984gg}} and is commonly used with great success for low-energy hadron phenomenology.
See Refs.~\cite{Scherer:2012xha,Pich:2018ltt} for modern introductions to ChPT.

There has been recent interest in the $3\to3$ meson scattering amplitude driven by advances in lattice QCD~\cite{Hansen:2014eka,Hansen:2015zga,Hansen:2019nir,Hammer:2017uqm,Hammer:2017kms,Mai:2017bge,Mai:2018djl,Mai:2019fba,Culver:2019vvu,Mai:2021lwb,Brett:2021wyd,Blanton:2019vdk,Blanton:2021llb}.
While many ChPT observables are known to high loop level, the six-meson amplitude was only recently calculated to one-loop level~\cite{Bijnens:2021hpq}, and then only for two quark flavors, i.e.\ a meson spectrum of only pions.
The case of three or more flavors is largely unexplored; the tree-level part is known up to next-to-next-to-next-to-leading order (N$^3$LO) in the massless case~\cite{Bijnens:2019eze}.
The leading-order (LO) massless pion case was initially done with current algebra methods and predates ChPT~\cite{Osborn:1969ku,Susskind:1970gf}.

While QCD is the canonical example, 
strongly coupled gauge theories can have different patterns of spontaneous symmetry breaking.
These were first discussed in the context of technicolor theories~\cite{Peskin:1980gc,Preskill:1980mz,Dimopoulos:1979sp}.
When the gauge group is vector-like and all fermions have the same mass, only three patterns show up as discussed in \rcite{Kogut:2000ek}; earlier work can be traced from there.
If all $n$ fermions are in a complex representation, the global symmetry group is $\SU(n)\times\SU(n)$ and is broken spontaneously to the diagonal (vector) $\SU(n)$, which corresponds to the $n$-quark QCD case.
If the fermions are in a real or pseudo-real representation, the global symmetry group is $\SU(2n)$ and is spontaneously broken to $\SO(2n)$ and $\Sp(2n)$, respectively.
We will refer to these cases as $\SU$, $\SO$ and $\Sp$, respectively, and collectively dub them `QCD-like theories'.
ChPT has been extended to these, and results can be found e.g.\ in \rrcite{Gasser:1986vb,Chivukula:1992gi,Splittorff:2001fy}.
The similarity between all cases, and a number of calculations to two-loop order (vacuum expectation value, mass and decay constant, and four-meson amplitudes), were worked out in \rrcite{Bijnens:2009qm,Bijnens:2011fm,Bijnens:2011xt}.

In the context of studying general properties of amplitudes, much attention has been paid to the structure of (massive) nonlinear sigma models 
(corresponding to ChPT without additional fields) 
at tree level including higher orders using various techniques~\cite{Low:2019ynd,Low:2020ubn,Kampf:2013vha,Kampf:2019mcd,Bijnens:2019eze}.
However, not all of these properties generalize to loop level.
Some loop-level progress can be found in \rcite{masterthesis,Farrow:2020voh,Bartsch:2022pyi}.
In this work, we calculate the six-meson amplitude to next-to-leading order (NLO) for the three symmetry-breaking patterns.
This generalizes the earlier work of \rcite{Bijnens:2021hpq} for the symmetry-breaking pattern $\OO(N+1)/\OO(N)$.

In \cref{sec:theory}, Chiral Perturbation Theory for QCD-like theories is shortly discussed; a more extensive discussion can be found in \rcite{Bijnens:2009qm}.
We introduce here also a notation that explicitly brings out the similarities for the three cases.
We do not describe the calculation in great detail;
it follows the standard Feynman diagram method and does sums over flavor indices using \cref{eq:fierz-SU,eq:fierz-SOp}.
The flavor structure of the general four- and six-meson amplitudes is discussed in Sec.~\hyperlink{sec:generalflavor}{III--A}.
The expressions can be very much simplified by using all symmetry properties of the amplitudes, as expected from general considerations.
This is described later in \cref{sec:ampl}.
We have checked that our results are UV finite and independent of the parametrization of the Nambu--Goldstone boson manifold, and that they reduce to the results of \rrcite{Bijnens:2011fm,Bijnens:2021hpq} in the appropriate cases.
At the end of \cref{sec:ampl}, we also present a limit of the six-meson amplitude in which the three-momenta of the particles of modulus $p$ go symmetrically to zero.
In this particular kinematic setting, we plot the flavor-stripped amplitudes with respect to $p$ and show the results in \cref{sec:numerics}.
Our conclusions are shortly discussed in \cref{sec:conclusions}, followed by several technical appendices that fix the notation and explain further subtleties and broader context.
Explicit expressions for our main result --- the NLO six-meson amplitude --- in terms of deorbited group-universal subamplitudes can be found in \cref{app:6pi-res}.

The analytical work in this manuscript was done both using Wolfram {\em Mathematica} with the \textsc{FeynCalc} package~\cite{Mertig:1990an,Shtabovenko:2016sxi,Shtabovenko:2020gxv} and a \textsc{FORM}~\cite{Vermaseren:2000nd} implementation.
The numerical results use \textsc{LoopTools}~\cite{vanOldenborgh:1989wn,Hahn:1998yk}.

%%%%%%%%%%%%%%%%%%%%%%%%%%%%%%%%%%%%%%%%%%%%%%%%%%%%%%%%%%%%%%%%%%%%%%%%%%%%%%%%
\section{Theoretical setting}
\label{sec:theory}

%%%%%%%%%%%%%%%%%%%%%%%%%%%%%%%%%%%%%%%%%%%%%%%%%%%%%%%%%%%%%%%%%%%%%%%%%%%%%%%%%
\subsection{Lagrangian}

We consider a theory of $n$ fermions with some symmetry group $G$, which is spontaneously broken to a subgroup $H$.
This gives rise to an EFT whose degrees of freedom are pseudo-Nambu--Goldstone bosons transforming under the quotient group $G/H$.
In analogy with the QCD case, we will refer to these as `mesons'.

We choose $G/H$ from the patterns of symmetry breaking present in the QCD-like theories described in the introduction.
The mesons are parametrized through a flavor-space matrix field $u$, also called the Nambu--Goldstone boson matrix.
In addition, the Lagrangian can be extended in terms of vector, axial-vector, scalar and pseudoscalar external fields~\cite{Gasser:1983yg,Gasser:1984gg}.
These correspond to vector and scalar sources for conserved and broken generators in general.
The symmetry may be explicitly broken by introducing quark masses in the scalar external field. Except for the definition of the decay constant and the introduction of quark masses, we do not need external fields in this work.

The Lagrangian for the meson--meson scattering at NLO relevant for all the discussed theories can be written as
\begin{equation}
\mathcal{L}
=\mathcal{L}^{(2)}+\mathcal{L}^{(4)}\,,
\label{eq:L}
\end{equation}
separating the LO and NLO terms in the chiral counting.%
\footnote{
    The next-to-next-to-leading-order (NNLO) terms  $\mathcal{L}^{(6)}$~\cite{Bijnens:1999sh} and N$^3$LO terms  $\mathcal{L}^{(8)}$~\cite{Bijnens:2018lez} are also known but not used here.}$^,$%
\footnote{
    Recall that the chiral counting order of an $\ell$-loop diagram with $n_k$ vertices from $\mathcal{L}^{(k)}$ is $m = 2 + 2\ell + \sum_k n_k(k-2)$.
    Thus, NLO ($m=4$) diagrams have either one loop or one vertex from $\mathcal{L}^{(4)}$.}
These take the form
\begin{align}
\mathcal{L}^{(2)}
&=\frac{F^2}4\tr{u_\mu u^\mu+\chi_+}\,,\\
\begin{split}
\mathcal{L}^{(4)}
&=L_0\tr{u_\mu u_\nu u^\mu u^\nu}
+L_1\tr{u_\mu u^\mu}\tr{u_\nu u^\nu}\\
&+L_2\tr{u_\mu u_\nu}\tr{u^\mu u^\nu}
+L_3\tr{u_\mu u^\mu u_\nu u^\nu}\\
&+L_4\tr{u_\mu u^\mu}\tr{\chi_+}
+L_5\tr{u_\mu u^\mu\chi_+}\\
&+L_6\tr{\chi_+}^2
+L_7\tr{\chi_-}^2
+\tfrac12L_8\tr{\chi_+^2+\chi_-^2}\,.
\end{split}
\label{eq:L4}
\end{align}
Above, $\tr{\cdots}$ denotes a flavor-space trace over $n\times n$ matrices for the $\SU$ and $2n\times 2n$ matrices for the $\SO$ and $\Sp$ cases.
Moreover,
\begin{align}
u_\mu&\equiv i\big(u^\dag\partial_\mu u-u\partial_\mu u^\dag\big)\,,\\
\chi_\pm&\equiv u^\dag\chi u^\dag\pm u\chi^\dag u\,.
\end{align}
Under $G$, both $u_\mu$ and $\chi_\pm$ transform as $X\to hXh^\dag$, where $h\in H$.
Above, as usual, $\chi\equiv2B_0\mathcal{M}$, with $\mathcal{M}=s-ip$, where $s(p)$ are the (pseudo)scalar external fields and $B_0$ is a parameter related to the scalar singlet quark condensate $\langle0|\bar qq|0\rangle$ (not to be confused with the integral $B_0$ in \cref{app:integrals}).
For our application and in the case with all the mesons having the same (lowest-order) mass $M$, we can simply put $\chi=M^2\mathbbb{1}$.

The Nambu--Goldstone boson
%flavor-space
matrix $u$ can be parametrized as
\begin{equation}
u=\text{exp}\bigg(\frac{i}{\sqrt{2}F}\,\phi^at^a\bigg)\,,
\label{eq:u_exp}
\end{equation}
where $\phi^a$ denote the pseudoscalar meson fields and $t^a$ are Hermitian generators of $G/H$ normalized to $\tr{t^at^b}=\delta^{ab}$.
Besides the `exponential' parametrization~\eqref{eq:u_exp}, there are other options available in the literature.
For practical calculations, it is useful to employ several different parametrizations in parallel.
This serves as a neat cross-check since, as anticipated, the final amplitude should be parametrization-independent.
We discuss the most general reparametrization in \cref{app:params}.
In the case of the six-meson amplitude at NLO, 18 free parameters appear in the expansion of $u$ in terms of $\phi^at^a$. We have checked that all our physical results are independent of these parameters.

%%%%%%%%%%%%%%%%%%%%%%%%%%%%%%%%%%%%%%%%%%%%%%%%%%%%%%%%%%%%%%%%%%%%%%%%%%%%%%%%%%%%%
\subsection{Flavor structures}

Each meson $\phi^a$ carries a \emph{flavor index} $a$, which appears in the amplitude carried by a $G/H$ generator residing in a flavor-space trace.
When a pair of fields is Wick contracted, the corresponding flavor indices are summed over; under $\SU$, the resulting expressions are evaluated using the Fierz identities
\begin{subequations}\label{eq:fierz-SU}
    \begin{alignat}{4}
        \tr{t^a A}\tr{t^a B}    &= \;\,\tr{AB} &&- \tfrac{1}{n}\tr{A}\tr{B}\,,\label{eq:fierz-SU:tree}\\
        \tr{t^a A t^a B}        &= \tr{A}\tr{B} &&- \tfrac{1}{n}\tr{AB}\,,\label{eq:fierz-SU:loop}
    \end{alignat}
\end{subequations}
where $A$ and $B$ are arbitrary flavor-space matrices.
The analogous identities for $\SO$ and $\Sp$ are quite similar, so we will use the abbreviation $\SOp$ in correlation with $\pm$: $\SO$ is paired with $+$, and $\Sp$ with $-$.
Thus, for $\SOp$ one uses
\begin{subequations}\label{eq:fierz-SOp}
    \begin{alignat}{6}
        \tr{t^a A}\tr{t^a B}    &= \tfrac12\big[&\tr{AB}\;\,&+\tr{AB^\R}\big] &&- \tfrac{1}{2n}\tr{A}\tr{B},\label{eq:fierz-SOp:tree}\\
        \tr{t^a A t^a B}        &= \tfrac12\big[&\tr{A}\tr{B}&\pm\tr{AB^\R}\big] &&- \tfrac{1}{2n}\tr{AB}\,.\label{eq:fierz-SOp:loop}
    \end{alignat}
\end{subequations}
Here, $B$ must be a string of generators $t^a$ or the unit matrix, so $\R$ effectively denotes reversal.%
\footnote{
    The general version is given in \rcite{Bijnens:2011fm}. The version here follows since the generators are Hermitian: $(t^at^b\cdots t^c)^\R = t^c\cdots t^bt^a$.}

Note that the implicitly summed index $a$ has different dimensions in \cref{eq:fierz-SU,eq:fierz-SOp}, corresponding to the number of mesons: $n^2-1$ under $\SU$ and $2n^2\pm n - 1$ under $\SOp$.
Note also, that due to the formally identical Lagrangians, \cref{eq:fierz-SU,eq:fierz-SOp} are the only source of formal dissimilarity between the amplitudes for the different cases.

\subsection{Low-energy constants and renormalization}
\label{sec:LECs}

At LO, we have two low-energy parameters: the mass $M$ (related to the aforementioned $B_0$) and decay constant $F$.
At NLO, 9 more constants (LECs) $L_i$ accompanying additional allowed chirally symmetric structures (operators) relevant for our application appear, as shown in \cref{eq:L4}.
These constants contain UV-divergent parts represented by coefficients $\Gamma_i$, which are uniquely fixed from the requirement that physical NLO amplitudes should be finite, and UV-finite parts $L_i^\r\equiv L_i^\r(\mu)$, renormalized at a scale $\mu$, that are free parameters in the theory:
\begin{equation}
L_i
=(c\mu)^{d-4}\left(\frac1{16\pi^2}\frac1{d-4}\,\Gamma_i+L_i^\r(\mu)\right).
\label{eq:li}
\end{equation}
Above, $d$ is the space-time dimension in the vicinity of 4 and $c$ is such that
\begin{equation}
\log c
=-\frac12\left(1-\gamma_\text{E}+\log4\pi\right).
\end{equation}
Consequently, in terms of $\epsilon=2-d/2$ and
\begin{equation}
\frac1{\tilde\epsilon}
\equiv\frac{1}{\epsilon}-\gamma_\text{E}+\log4\pi-\log\mu^2+1\,,
\label{eq:epstilde}
\end{equation}
one writes (to NLO)
\begin{equation}
L_i
=-\kappa\,\frac{\Gamma_i}2\frac1{\tilde\epsilon}+L_i^\r\,,
\label{eq:li2}
\end{equation}
with $\kappa\equiv\frac1{16\pi^2}$.
The extra `+1' term in \cref{eq:epstilde} with respect to the standard $\overline{\text{MS}}$ renormalization scheme is customary in ChPT.

Studying and renormalizing the four-meson amplitude at NLO (i.e.\ considering one-loop diagrams with vertices from $\mathcal{L}^{(2)}$ and tree-level counterterms from $\mathcal{L}^{(4)}$) determines all the $\Gamma_i$ except for one:
The divergent part of $L_7$ remains unset.
It can, however, be fixed from the six-meson amplitude.
Using the heat-kernel technique, all NLO divergences were derived in \rcite{Bijnens:2009qm}.
For the reader's convenience, we list the $\Gamma_i$ here in a group-universal form:
\begin{alignat}{4}
\label{eq:divergence}
        \Gamma_0 &= \tfrac1{48}(n+4\sgn)\,,\qquad & \Gamma_5 &= \tfrac n8\,, \notag\\
        \Gamma_1 &= \tfrac1{16\kk}\,, & \Gamma_6 &= \tfrac1{16\kk}+\tfrac{1}{8\kk^2n^2}\,, \notag\\
        \Gamma_2 = \Gamma_4 &= \tfrac1{8\kk}\,, & \Gamma_7 &= 0\,, \notag\\
        \Gamma_3 &= \tfrac1{24}(n-2\sgn)\,, & \Gamma_8 &= \tfrac1{16}(n+\sgn-\tfrac{4}{\kk n})\,.
\end{alignat}
Above, $\sgn$ and $\kk\equiv(1+\sgnp)$ parametrize the groups as follows:
\begin{equation}\label{eq:sgn-kk}
    % \kk\equiv(1+\sgnp),\qquad
    \sgn\equiv 
    \begin{cases}
        \hphantom{\pm}0 & \big[\SU\big]\,,\\
        \pm1 & \big[\SOp\big]\,,   
    \end{cases}
    \qquad
    \kk= 
    \begin{cases}
        1 & \big[\SU\big]\,,\\
        2 & \big[\SOp\big]\,.   
    \end{cases}
\end{equation}
Another check on our calculation is that, with the expressions in \cref{eq:divergence}, all our results are finite.

\subsection{Mass and decay constant}
\label{sec:massdecay}

The $Z$ factor used for the wave-function renormalization is related to the meson self-energy $\Sigma$ as
\begin{equation}
\frac1Z
=1-\frac{\partial\Sigma(p^2)}{\partial p^2}\bigg|_{p^2=M_\pi^2}\,,
\end{equation}
with $-i\Sigma$ being represented by a tadpole graph with two external legs plus counterterms stemming from the Lagrangian~\eqref{eq:L4}.
Note that in our application the physical mass of all mesons is equal and denoted as $M_\pi$.
At NLO, the LO vertex and propagator are extended in terms of the replacements
\begin{equation}
\begin{split}
M^k&\to M_\pi^k\bigg(1-\frac k2\frac{\overline\Sigma}{M_\pi^2}\bigg)\,,\\
\frac1{F^k}&\to\frac1{F_\pi^k}(1+k\delta F)\,,
\end{split}
\end{equation}
at the given order equivalent to the standard $M_\pi^2=M^2+\overline\Sigma$, $F_\pi=F(1+\delta F)$, with
\begin{equation}
\label{eq:massdecay}
\begin{split}
\overline\Sigma
&=\frac{M_\pi^4}{F_\pi^2}
\biggl\{
-8\big[L_5^\r-2L_8^\r+n\kk(L_4^\r-2L_6^\r)\big]\\
&\qquad\qquad+\bigg(\frac1{\kk n}-\frac\sgn2\bigg)L
\biggr\}
+\mathcal{O}\bigg(\frac1{F_\pi^4}\bigg)\,,\\
\delta F
&=\frac{M_\pi^2}{F_\pi^2}
\Big[
4(L_5^\r+n\kk L_4^\r)
-\frac n2L
\Big]
+\mathcal{O}\bigg(\frac1{F_\pi^4}\bigg)\,.
\end{split}
\end{equation}
Here, we again present the group-universal form.
Above and later on, we use $L\equiv\kappa\log\frac{M^2}{\mu^2}$.
Needless to say, in the final result one only retains the terms relevant at order $\mathcal{O}(p^4)$.
Thus, in the rest of the NLO expressions one simply takes $M\to M_\pi$ and $F\to F_\pi$.
Note that we recalculated the results of \cref{eq:massdecay} and that they agree with \rrcite{Gasser:1986vb,Splittorff:2001fy,Bijnens:2009qm}.

%%%%%%%%%%%%%%%%%%%%%%%%%%%%%%%%%%%%%%%%%%%%%%%%%%%%%%%%%%%%%%%%%%%%%%%%%%%%%%%%%
\section{The amplitudes}
\label{sec:ampl}

In terms of Feynman diagrams, loop integrals, etc., the present calculation proceeds along the same lines as the one performed in \rcite{Bijnens:2021hpq}.
However, the result is considerably more cumbersome, largely because the flavor indices are carried by more structures beyond Kronecker $\delta$'s.
There is also the matter of treating the $\SU$, $\SO$ and $\Sp$ variants in parallel without tripling the amount of material to present.
We will therefore devote much of this section to simplifying the amplitude expressions.

%%%%%%%%%%%%%%%%%%%%%%%%%%%%%%%%%%%%%%%%%%%%%%%%%%%%%%%%%%%%%%%%%%%%%%%%%%%%%%%%%%%%%
\subsection{Flavor structure of the four- and six-meson amplitudes}
\hypertarget{sec:generalflavor}{}

The general flavor structure of the $\SU$ case for meson--meson scattering is well-known; see e.g.\ \rrcite{Chivukula:1992gi,Bijnens:2009qm}.
We consider four incoming mesons with flavor indices $\f_1,\ldots,\f_4$ and momenta $p_1,\ldots,p_4$.
The usual Mandelstam variables are defined as%
\footnote{We have chosen the specific definition here to later define an off-shell extension.}
\begin{equation}\label{eq:mandel}
    s = \left(p_1+p_2\right)^2,\, t=\left(p_1+p_3\right)^2,\, u=\left(p_2+p_3\right)^2.
\end{equation}
The amplitude is then conventionally decomposed as
\begin{align}\label{eq:BC}
    \ampl_{4\pi}(s,t,u) &= \big(\tr{t^{\f_1} t^{\f_2} t^{\f_3} t^{\f_4} }+\tr{t^{\f_4} t^{\f_3} t^{\f_2} t^{\f_1} }\big) B(s,t,u)\notag\\
    &+\big(\tr{t^{\f_1} t^{\f_3} t^{\f_4} t^{\f_2} }+\tr{t^{\f_2} t^{\f_4} t^{\f_3} t^{\f_1} }\big) B(t,u,s)\notag\\
    &+\big(\tr{t^{\f_1} t^{\f_4} t^{\f_2} t^{\f_3} }+\tr{ t^{\f_3} t^{\f_2} t^{\f_4} t^{\f_1} }\big) B(u,s,t)\notag\\
    &+\delta^{\f_1 \f_2} \delta^{\f_3 \f_4}C(s,t,u)+\delta^{\f_1 \f_3}\delta^{\f_2 \f_4} C(t,u,s)\notag\\
    &+\delta^{\f_1 \f_4}\delta^{\f_2 \f_3} C(u,s,t)\,.
\end{align}
The functions satisfy $B(s,t,u) = B(u,t,s)$ and $C(s,t,u)=C(s,u,t)$. 
This structure follows from requiring invariance under the unbroken group, Bose symmetry and charge conjugation for $\SU$. 
Under $\SOp$, $\tr{t^at^bt^ct^d}=\tr{t^dt^ct^bt^a}$ without relying on charge conjugation.

A similar decomposition for the six-meson amplitude in terms of six flavor labels and momenta is
\begin{align}\label{eq:DEFG}
    \ampl_{6\pi}(p_1,.\,.&\,.\,,p_6)\notag\\
    =\sum_{\mathcal S_6}\bigg\{
            &\tfrac1{12}\pig[\tr{t^{\f_1}\cdots t^{\f_6}}+\tr{t^{\f_6}\cdots t^{\f_1}}\pig] D(p_1,\ldots,p_6)\notag\\
            +&\tfrac1{16} \delta^{\f_1\f_2}\pig[\tr{t^{\f_3}\cdots t^{\f_6}}+\tr{t^{\f_6}\cdots t^{\f_3}}\pig] E(p_1,\ldots,p_6)\notag\\
            +&\tfrac1{36}\pig[\tr{t^{\f_1}t^{\f_2}t^{\f_3}}\tr{t^{\f_4}t^{\f_5}t^{\f_6}}\notag\\
            &\qquad+\tr{t^{\f_3}t^{\f_2}t^{\f_1}}\tr{t^{\f_6}t^{\f_5}t^{\f_4}}\pig] F(p_1,\ldots,p_6)\notag\\
            +&\tfrac1{48}\delta^{\f_1 \f_2} \delta^{\f_3 \f_4}\delta^{\f_5 \f_6}G(p_1,\ldots,p_6) \bigg\}\,,
\end{align}
where $\mathcal S_6$ represents the $6!=720$ permutations of $\{1,\ldots,6\}$ and the symmetry factors correspond to how many permutations leave the traces and $\delta$'s (the `flavor structure') unchanged; thus, $D$, $E$, $F$ and $G$ are summed over 60, 45, 20 and 15 distinct permutations, respectively, just like $B$ and $C$ are summed over 3.
Charge conjugation and group structure imply the following properties:
\begin{subequations}\label{eq:DEFG-sym}
    \begin{align}
        D(p_1,\ldots,p_6)&= D(p_2,p_3,p_4,p_5,p_6,p_1)\notag\\
            &= D(p_6,p_5,p_4,p_3,p_2,p_1)\,,\\
        E(p_1,\ldots,p_6) &= E(p_2,p_1,p_3,p_4,p_5,p_6)\notag\\
            &=E(p_1,p_2,p_4,p_5,p_6,p_3)\notag\\
            &=E(p_1,p_2,p_6,p_5,p_4,p_3)\,,\\
        F(p_1,\ldots,p_6)&=F(p_4,p_5,p_6,p_1,p_2,p_3)\notag\\
            &=F(p_2,p_3,p_1,p_4,p_5,p_6)\notag\\
            % &=F(p_1,p_2,p_3,p_5,p_6,p_4)\notag\\
            &=F(p_1,p_3,p_2,p_4,p_6,p_5)\,,\\
        G(p_1,\ldots,p_6)&=G(p_2,p_1,p_3,p_4,p_5,p_6)\notag\\
            &=G(p_3,p_4,p_5,p_6,p_1,p_2)\notag\\
            &=G(p_3,p_4,p_1,p_2,p_5,p_6)\,.
    \end{align}
\end{subequations}
These are discussed in a more formal and general way in the next subsection.

%%%%%%%%%%%%%%%%%%%%%%%%%%%%%%%%%%%%%%%%%%%%%%%%%%%%%%%%%%%%%%%%%%%%%%%%%%%%%%%%%%%%%
\subsection{General flavor-based simplification}
\label{sec:strip}

In order to formalize the structure seen in \cref{eq:BC,eq:DEFG}, we follow the notation of \rcite{Bijnens:2019eze} and define a $\N$-particle \emph{flavor structure} as
\begin{multline}
    \flav_R(\f_1,\ldots,\f_\N) = \tr{t^{\f_{1}}t^{\f_{2}}\cdots t^{\f_{r_1}}}\\\times\tr{t^{\f_{r_1+1}}\cdots t^{\f_{r_1+r_2}}}\cdots\tr{t^{\f_{\N-r_{|R|}}+1}\cdots t^{\f_{\N}}}\,,
\end{multline}
where $R=\{r_1,\ldots,r_{|R|}\}$ with $\sum r_i = \N$ is a \mbox{\emph{flavor split}}: The flavors are split across $|R|$ traces, each containing $r_i$ indices.
Without loss of generality, we may impose $r_1\leq r_2\leq\cdots \leq r_{|R|}$.
For a permutation $\sigma$ that maps $i\to\sigma_i$, we define $\flav_R^\sigma(\f_1,\ldots,\f_\N)\equiv\flav_R(\f_{\sigma_{1}},\ldots,\f_{\sigma_{\N}})$ and denote by $\Z_R$ the group of permutations that preserve $\flav_R$: For all $\sigma\in\Z_R$, $\flav_R^\sigma(\f_1,\ldots,\f_\N)=\flav_R(\f_1,\ldots,\f_\N)$.%
\footnote{
    $\Z_R$ is the cyclic group $\Z_\N$ when $R=\{\N\}$, hence the notation.
    In general, it combines cyclic symmetry of each trace with exchanging the contents of same-size traces.
    It is Abelian as long as all $r_i$ are different.}
The group $\Z_R$ is, of course, related to the symmetries in \cref{eq:DEFG-sym}.
    
In general, an amplitude can be decomposed as
\begin{multline}\label{eq:strip}
    \ampl_{k\pi}(p_1,\f_1;p_2,\f_2;\ldots;p_\N,\f_\N) \\= \sum_R \sum_{\sigma} \ampl^\sigma_R(p_1,\ldots,p_\N)\flav_R^\sigma(\f_1,\ldots,\f_\N)\,,
\end{multline}
where $\sigma$ is summed over all permutations that do not preserve $\flav_R$, i.e.\ $\mathcal S_\N/\Z_R$.
It follows from Bose symmetry that $\ampl^\sigma_R(p_1,\ldots,p_\N) = \ampl^{\id}_R(p_{\sigma_{1}},\ldots,p_{\sigma_{\N}})$ where $\id$ is the identity permutation. 
It is therefore sufficient to work with $\ampl_R\equiv\ampl^{\id}_R$, the \emph{stripped amplitude}, for all $R$; the full amplitude follows from \cref{eq:strip}.

The stripped amplitude is easily obtained from the full amplitude by taking the coefficient of $\flav_R$.
In $\SU$, it is guaranteed to be unique, as was proven in \rcite{Bijnens:2019eze}.
This carries over to $\SO$ and $\Sp$; the ambiguity created by $\tr{X}=\tr{X^\R}$ is easily resolved by applying $\tr{X}\to\frac12[\tr{X}+\tr{X^\R}]$ before extracting $\ampl_R$.

In a four-meson amplitude, the stripped amplitudes $\sampl{4}$ and $\sampl{2,2}$ are the functions called $B(s,t,u)$ and $C(s,t,u)$, respectively, in \cref{eq:BC}.
For six mesons, one has $\sampl{6}$, $\sampl{2,4}$, $\sampl{3,3}$ and $\sampl{2,2,2}$, which correspond to $D$, $E$, $F$ and $G$ in \cref{eq:DEFG}, respectively.
In the $\SU(n=2)$ case (equivalent to the $\OO(4)/\OO(3)$ case treated in \rcite{Bijnens:2021hpq}), the Cayley--Hamilton theorem allows all the trace structures to be reduced to $R=\{2,2,2\}$.
When $n=2,3,4,5$ for $\SU$ and $n=1,2$ for $\SOp$, respectively, the $\flav_R$ satisfy a number of linear relations (see \rcite{Zhang:2007qp} for explicit expressions), which in turn relate the $\ampl_R$ to each other.
Otherwise, $\flav_R$ are linearly independent for different $R$.%
\footnote{
    They are in fact orthogonal in a certain sense, as shown in \rcite{Bijnens:2019eze}.}

As follows from its definition, $\ampl_R$ inherits $\Z_R$ symmetry (acting on $\{p_1,\ldots,p_\N\}$) from $\flav_R$.
We must also consider another permutation of the external particles, which we dub \emph{trace-reversal} (TR).
It is the permutation which reverses the product of generators in each trace: $\tr{t^at^b\cdots t^c}\to\tr{t^c\cdots t^bt^a}$.
Under $\SU$, this is \emph{not} a symmetry of $\flav_R$, but CP invariance nevertheless requires it to be a symmetry of $\ampl_R$: Charge conjugation maps $t^a\to(t^a)^{\!\T}$, and thus $\tr{t^at^b\cdots t^c}\to\tr{t^{a\T}t^{b\T}\cdots t^{c\T}}=\tr{t^c\cdots t^bt^a}$.
This is why \cref{eq:BC,eq:DEFG} pair each trace with its reverse (except for the reversal-symmetric $\tr{t^at^b}$).
We will denote the general symmetry group of $A_R$, i.e.\ $\Z_R$ plus TR, by $\Z_R^\rev$.

Under $\SOp$, $\Z_R^\rev$ is a symmetry also of $\flav_R$; in fact, $\tr{t^at^b\cdots t^c}=\tr{t^c\cdots t^bt^a}$ makes $\flav_R$ symmetric under the reversal of any single trace (CP only requires symmetry under the simultaneous reversal of all traces).
This enhanced symmetry is inherited by $A_R$, and is very important for the relation between the amplitudes of the different QCD-like theories (see \cref{app:groups}).

The size of the amplitude expressions can be further reduced by writing them in terms of a quantity $\dampl_R$ such that
\begin{equation}\label{eq:deorbit}
    \ampl_R(p_1,\ldots,p_\N) = \sum_{\sigma\in\Z^\rev_R} \dampl_R(p_{\sigma_{1}},\ldots,p_{\sigma_{\N}})\,.
\end{equation}
This clearly exists (consider e.g.\ $\dampl_R=\ampl_R/|\Z^\rev_R|$) but is not unique.
A method for obtaining a minimal-length $\dampl_R$, the \emph{deorbited} stripped amplitude, is described in \cref{app:deorbit}.

%%%%%%%%%%%%%%%%%%%%%%%%%%%%%%%%%%%%%%%%%%%%%%%%%%%%%%%%%%%%%%%%%%%%%%%%%%%%%%%%
\subsection{Group-universal formulation}\label{sec:group-univ}

One can expect the amplitudes of the $\SU$, $\SO$ and $\Sp$ theories to have many similarities, since the only differences relevant to the amplitude are the variations of the Fierz identity, \cref{eq:fierz-SU,eq:fierz-SOp}, and the substitution $n\to\kk n$.
In fact, comparison of the amplitudes suggests that one might introduce four subamplitudes $\gampl{i}$ such that
\begin{equation}\label{eq:group-univ}
    \ampl = \bigg\{\gampl{1} + \sgn\gampl{\sgn} + \sgnp\gampl{\sgnp} + \frac{\gampl{\kk}}{\kk}\bigg\}_{\stackrel{}{n\to\kk n}}\,,
\end{equation}
where $(\sgn,\kk)=(0,1)$ for $\SU$ and $(\pm1,2)$ for $\SOp$, as defined in \cref{eq:sgn-kk}.
This decomposition is clearly redundant: Three amplitudes are expressed as a combination of four subamplitudes.
However, we find it natural and choose it for its simplicity and clarity; very few terms appear in more than one subamplitude, and $\gampl{\sgnp}$ is a relatively short expression.
The decomposition~\eqref{eq:group-univ} can be combined with stripping and deorbiting, allowing the amplitude to be formulated using the very concise quantities $\dgampl{i}_R$.
Furthermore, many of these are actually zero.
The patterns for which $(R,i)$ combinations are allowed, what LECs, loop integrals and powers of $n$ may appear where, etc., are studied in \cref{app:6pi-res} and explained in \cref{app:groups}.

%%%%%%%%%%%%%%%%%%%%%%%%%%%%%%%%%%%%%%%%%%%%%%%%%%%%%%%%%%%%%%%%%%%%%%%%%%%%%%%%%%%%
\subsection{The four-meson amplitude}

\begin{figure}[t]
    \centering
    \includegraphics{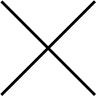}
    \caption{
        The single LO four-meson diagram, with the vertex stemming from $\lagr^{(2)}$.
        In formulae we refer to it as $i\ograph{LO}{2}$ or, after NLO mass and decay-constant redefinitions~\eqref{eq:massdecay} are applied, $i\ograph{LO}{4}$.}
    \label{fig:4piLO}
\end{figure}

\begin{figure}[t]
    \centering
        \begin{subfigure}[t]{0.32\columnwidth}
            \includegraphics{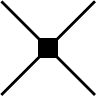}
            \caption{$1\times$}
            \label{fig:4piNLO:point}
        \end{subfigure}
        \begin{subfigure}[t]{0.32\columnwidth}
            \includegraphics{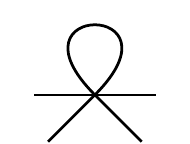}
            \caption{$1\times$}
            \label{fig:4piNLO:A}
        \end{subfigure}
        \begin{subfigure}[t]{0.32\columnwidth}
            \includegraphics{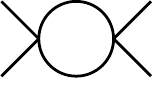}
            \caption{$3\times$}
            \label{fig:4piNLO:B}
        \end{subfigure}
    \caption{
        NLO topologies for the four-meson amplitude.
        The unmarked vertices stem from $\lagr^{(2)}$, while the square vertices stem from $\lagr^{(4)}$ and contain $L_i$, $i=1,\ldots,8$.
        The numbers (diagram multiplicities) indicate the number of distinct diagrams with the same topology but different permutations of the external legs.
        Diagram (a) corresponds to $i\graph{CT}$ and diagrams (b,c) sum up to $i\graph{1-loop}$.}
    \label{fig:4piNLO}
\end{figure}

The notation of the previous sections allows the four-meson amplitude to be written very compactly.
We will use the ordinary Mandelstam variables~\eqref{eq:mandel}.
At LO, there is a single nonzero subamplitude, stemming from the diagram in \cref{fig:4piLO},%
\footnote{
    We remind the reader that the $R=\{\ldots\}$ subscript indicates stripping according to \cref{eq:strip}, the tilde indicates deorbiting according to \cref{eq:deorbit}, and the calligraphic $\mathcal A$ indicates group-universal formulation according to \cref{eq:group-univ}, with the subamplitude label `1' sharing the superscript with the `LO' label.
    All three simplifications can be applied independently and commute with each other.}
\begin{equation}
    \sgampl{4}{\text{LO},1} = 8\dsgampl{4}{\LO,1} = -\frac{t - 2M_\pi^2}{2F_\pi^2}\,.
\label{eq:A4pi_LO}
\end{equation}
At NLO, one has one-loop diagrams (two topologies of four one-loop diagrams in total) combined with counterterms, as shown in \cref{fig:4piNLO}.
Moreover, one needs to take into account NLO wave-function renormalization $(Z^{1/2}-1)$ applied for every external leg, and mass and decay-constant redefinitions [at the given order based on \cref{eq:massdecay}] applied to the LO graph $\ograph{LO}{2}$, giving $\ograph{LO}{4}$.
Schematically, this can be summed up as
\begin{equation}
A_{4\pi}^{(\NLO)}
=\graph{1-loop}
+\graph{CT}
+4(Z^{1/2}-1)\ograph{LO}{2}
+\ograph{LO}{4}\,.
\label{eq:A4pi_NLO}
\end{equation}
Note that while the above combination is parametri\-zation-independent and UV finite, the separate terms are not.
Altogether, the nonzero stripped and deorbited group-universal NLO subamplitudes read
\begin{subequations}\label{eq:4meson}
\begin{align}
    F_\pi^4 \dsgampl{4}{\NLO, 1} &= 
        \smallfrac{M_\pi^4}{4 n}  \pig\{L+\kappa - \bar J( s ) \pig\}    % M4p4R4.A1N.000.hf
        \notag\\
        &+  2 M_\pi^4 (L_0^\r+L_8^\r) 
            -  \smallfrac{M_\pi^2 t}{2} (4L_0^\r + L_5^\r) \notag\\
        &+ \smallfrac{t^2}{8} (4L_0^\r + L_3^r) 
            + \smallfrac{s(s-u)}{4} L_3^\r\,,     % M4p4R4.A1N.001.hf
    \\
    F_\pi^4 \dsgampl{4}{\NLO,\xi} &= 
        \smallfrac{ \bar J( t )}{64}    % M4p4R4.A2N.000.hf
        \pig\{
            (t-2 M_\pi^2)^2
        \pig\}\notag\\
        &+ \smallfrac{ \bar J( s )}{192}    % M4p4R4.A2N.001.hf
        \pig\{
            (4 M_\pi^2 -  s ) (3 s + t - u)
        \pig\}\notag\\
        &- \smallfrac{6L+5\kappa}{1152}    % M4p4R4.A2N.002.hf
        \pig\{
            28 M_\pi^4 - 16 M_\pi^2 t + 3 t^2  \eqbreak{-} 2 s (s-u)
        \pig\}
        - \smallfrac{ M_\pi^4 \kappa }{96} \,,
    \\
    F_\pi^4 \dsgampl{4}{\NLO,\zeta} &= 
        \smallfrac{n \bar J( s )}{192}      % M4p4R4.A3N.000.hf
        \pig\{
            4 M_\pi^2(t-u) + s(3 s - t + u)
        \pig\}\notag\\
        &- \smallfrac{n( 3L+2\kappa)}{1152}    % M4p4R4.A3N.001.hf
        \pig\{
            32 M_\pi^4 - 20 M_\pi^2 t + 3 t^2 \eqbreak{+} 2 s (s-u)
        \pig\} \notag\\
        &- \smallfrac{ n \kappa}{288}
        \pig\{
            4M_\pi^4 - M_\pi^2 t + s (s-u)
        \pig\}\,,    % M4p4R4.A3N.002.hf
    \\
    F_\pi^4 \dsgampl{2,2}{\NLO, 1} &= 
        \smallfrac{M_\pi^4}{4 n^2}  \pig\{ \bar J( s ) - (L+\kappa)\pig\}    % M4p4R22.A1N.000.hf
        + \smallfrac{u(u-t)}{2} L_2^\r\notag\\
        &+  4 M_\pi^4 (L_1^\r - L_4^\r + L_6^\r)
        + \smallfrac{s^2}{4}(4L_1^\r + L_2^\r) \notag\\     % M4p4R22.A1N.001.hf
        &-  2 M_\pi^2 s (2L_1^\r - L_4^\r)
        \,,    % M4p4R22.A1N.003.hf
    \\
    F_\pi^4 \dsgampl{2,2}{\NLO,\zeta} &= 
        \smallfrac{ s^2\bar J( s )}{32}
        + \smallfrac{ \bar J( u )}{16}    % M4p4R22.A3N.000.hf
        \pig\{
            (  u -2 M_\pi^2 )^2
        \pig\}\notag\\
        &- \smallfrac{ L+\kappa}{64}    % M4p4R22.A3N.002.hf
        \pig\{
          3 s^2 - 2 u (t-u)
        \pig\}\,.
\end{align}%
\end{subequations}
(Recall $\kappa$ and $L$ from \cref{sec:theory}; $\bar J$ is defined in \cref{app:integrals}.)
This is identical to the results given in \rcite{Bijnens:2011fm}.

%%%%%%%%%%%%%%%%%%%%%%%%%%%%%%%%%%%%%%%%%%%%%%%%%%%%%%%%%%%%%%%%%%%%%%%%%%%%%%%%%%%%%%%
\subsection{Poles and factorization}\label{sec:pole}

The six-meson amplitude has a simple pole whenever an internal propagator goes on-shell, i.e.\ $p_{ijk}^2 = M_\pi^2$ with $p_{ijk}=p_i+p_j+p_k$ for any indices $i,j$ and $k$.
As in \rcite{Bijnens:2021hpq}, the amplitude can therefore be separated into a part containing the pole and a nonpole part,
\begin{equation}\label{eq:pole-nonpole}
    \ampl_{6\pi} = \ampl_{6\pi}^{(\pole)} + \ampl_{6\pi}^{(\nonpole)}\,,
\end{equation}
where the pole part can be factorized in terms of four-meson amplitudes:
\begin{multline}\label{eq:factor}
    \ampl_{6\pi}^{(\pole)} = \sum_{P_{10}, \f_\text{o}}
        \ampl_{4\pi}(p_i,\f_i;p_j,\f_j;p_k,\f_k; -p_{ijk},\f_\text{o})\\
        \times \frac{-1}{p_{ijk}^2 - M^2_\pi}\,
        \ampl_{4\pi}(p_\ell,\f_\ell;p_m,\f_m;p_n,\f_n;p_{ijk},\f_\text{o})\,.
\end{multline}
Above, $P_{10}$ represents the 10 distinct ways of distributing the indices $1,\ldots,6$ into two triples $i,j,k$ and $\ell,m,n$, and $\f_\text{o}$ is the flavor of the off-shell leg, i.e.\ the propagator.

This factorization can also be done at the stripped-amplitude level.
With \cref{eq:factor} schematically summarized as $\ampl_{6\pi}^{(\pole)}\sim \ampl_{4\pi}\times \ampl_{4\pi}$, we correspondingly have
\begin{equation}
    \begin{aligned}
        2\left(\sampl{6}^{(\pole)} - \tfrac1n\sampl{3,3}^{(\pole)}\right) &\sim \sampl{4}\times\sampl{4}\,,\\
        \sampl{2,4}^{(\pole)} &\sim \sampl{2,2}\times\sampl{4}\,,\\
        2\sampl{2,2,2}^{(\pole)} &\sim \sampl{2,2}\times\sampl{2,2}\,,
    \end{aligned}
\end{equation}
with each $\ampl_R^{(\pole)}$ summed over $\Z_R$ instead of $P_{10}$, causing some symmetry factors.%
\footnote{
    Note that $\sampl{2,4}^{(\pole)}=0$ at LO and $\sampl{2,2,2}^{(\pole)}=0$ at NLO, since $\sampl{2,2}=0$ at LO.}

In \cref{eq:factor}, the four-pion subamplitude is defined as usual, although $s+t+u=3M_\pi^2+p_{ijk}^2$ since one leg is off-shell.
The residue at the pole is unique (since the on-shell four-meson amplitude is), but the extrapolation away from $p_{ijk}^2=M_\pi^2$ is not.
Correspondingly, the distribution of terms between the parts in \cref{eq:pole-nonpole} is not unique.
We choose to express $\ampl_{4\pi}$ in \emph{exactly} the form~\eqref{eq:4meson}, which in turn fixes $\ampl_{6\pi}^{(\nonpole)}$.
This choice by definition leaves both $\ampl_{6\pi}^{(\pole)}$ and $\ampl_{6\pi}^{(\nonpole)}$ parametrization-independent.
However, the distribution of contributions from individual one-particle-reducible (1PR) diagrams remains parametrization-dependent, while  one-particle-irreducible (1PI) diagrams only contribute to $\ampl_{6\pi}^{(\nonpole)}$.

By suitably deforming $A_{4\pi}$, it is in fact possible to make the tree-level $\ampl_{6\pi}^{(\nonpole)}$ vanish.
This is the principle underlying Britto--Cachazo--Feng--Witten recursion~\cite{Britto:2004ap,Britto:2005fq} and similar techniques, wherein many-particle amplitudes are recursively built up from smaller ones.
This technique was used for the first published calculation of the NLO tree-level six-meson amplitude~\cite{Low:2019ynd}, but at least its standard configuration suffers from convergence problems at NNLO~\cite{Bijnens:2019eze}.
Significant work has been done on the topic of loop-level recursion techniques but is typically limited to loop integrands rather than complete amplitudes; see \rcite{Farrow:2020voh,Bartsch:2022pyi} and references therein. 
We make no use of such techniques here.

%%%%%%%%%%%%%%%%%%%%%%%%%%%%%%%%%%%%%%%%%%%%%%%%%%%%%%%%%%%%%%%%%%%%%%%%%%%%%%%%%%%%%

\subsection{The six-meson amplitude}

\begin{figure}[t]
    \centering
    \begin{subfigure}[t]{0.45\columnwidth}
        \includegraphics{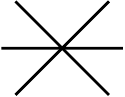}
        \caption{$1\times$}
        \label{fig:6piLO:point}
    \end{subfigure}
    \begin{subfigure}[t]{0.45\columnwidth}
        \includegraphics{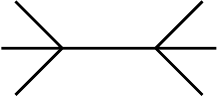}
        \caption{$10\times$}
        \label{fig:6piLO:prop}
    \end{subfigure}
    \caption{
        LO topologies for the six-meson amplitude.
        As in \cref{fig:4piNLO}, multiplicities are indicated.}
    \label{fig:6piLO}
\end{figure}

\begin{figure}[t]
    \centering
    \begin{subfigure}[t]{0.32\columnwidth}
        \includegraphics{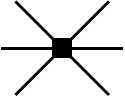}
        \caption{$1\times$}
        \label{fig:6piNLO:point}
    \end{subfigure}
    \begin{subfigure}[t]{0.32\columnwidth}
        \includegraphics{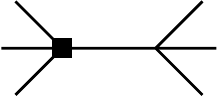}
        \caption{$20\times$}
        \label{fig:6piNLO:prop}
    \end{subfigure}
    \begin{subfigure}[t]{0.32\columnwidth}
        \includegraphics{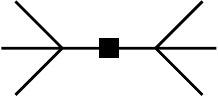}
        \caption{$10\times$}
        \label{fig:6piNLO:2prop}
    \end{subfigure}
    
    \begin{subfigure}[t]{0.32\columnwidth}
        \includegraphics{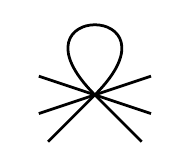}
        \caption{$1\times$}
        \label{fig:6piNLO:A-point}
    \end{subfigure}
    \begin{subfigure}[t]{0.32\columnwidth}
        \hskip-3mm  
        \includegraphics{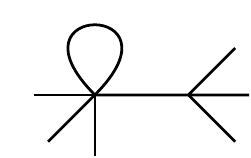}
        \caption{$20\times$}
        \label{fig:6piNLO:A-prop}
    \end{subfigure}
    \begin{subfigure}[t]{0.32\columnwidth}
        \includegraphics{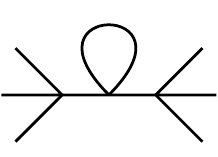}
        \caption{$10\times$}
        \label{fig:6piNLO:A-2prop}
    \end{subfigure}
    
    \begin{subfigure}[t]{0.32\columnwidth}
        \includegraphics{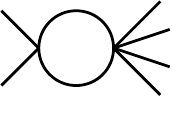}
        \caption{$15\times$}
        \label{fig:6piNLO:B-point}
    \end{subfigure}
    \begin{subfigure}[t]{0.32\columnwidth}
        \includegraphics{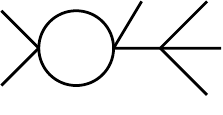}
        \caption{$60\times$}
        \label{fig:6piNLO:B-prop}
    \end{subfigure}
    \begin{subfigure}[t]{0.32\columnwidth}
        \includegraphics{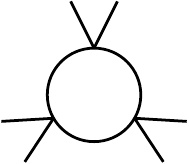}
        \caption{$15\times$}
        \label{fig:6piNLO:C}
    \end{subfigure}
    \caption{
        NLO topologies for the six-meson amplitude.
        As in \cref{fig:4piNLO}, multiplicities are indicated.}
    \label{fig:6piNLO}
\end{figure}

The nonpole LO six-meson amplitude contains only two nonzero subamplitudes, stemming from the 1 + 10 tree diagrams in \cref{fig:6piLO}:
\begin{subequations}
    \label{eq:6meson-LO}
    \begin{align}
        \sgampl{6}{\LO,\,\nonpole,\,1} &= \frac{p_1\cdot p_3 + p_3\cdot p_5 + p_5\cdot p_1 }{2F_\pi^4}\,,\\
        \sgampl{3,3}{\LO,\,\nonpole,\,1} &= \frac{M_\pi^2 -  p_1\cdot p_2 - p_1\cdot p_3 - p_2\cdot p_3}{2n F_\pi^4}\,.
    \end{align}
\end{subequations}
The pole part is given by \cref{eq:factor}.
Note that \cref{eq:6meson-LO} is group-invariant, i.e.\ equal for $\SU$ and $\SOp$ up to $n\to\kk n$.
This is true for all analogous LO $\N$-meson amplitudes, and indeed all tree-level contributions, as is proven in \cref{app:groups}.

The NLO amplitude, which is the main result of this work, stems from the diagrams in \cref{fig:6piNLO}.
Even when maximally simplified, it is rather lengthy, so we leave its explicit expressions to \cref{app:6pi-res}.

Let us now describe briefly the renormalization procedure for the NLO six-meson amplitude analogously to \cref{eq:A4pi_NLO}.
Regarding the 1PI diagrams [\cref{fig:6piNLO:point,fig:6piNLO:A-point,fig:6piNLO:B-point,fig:6piNLO:C}] which only contribute to the nonpole part, we can again write, schematically,
\begin{equation}
\begin{split}
A_{6\pi}^{(\NLO)}\big|_\text{1PI}
&={\mathcal{M}}_\text{1-loop,1PI}^{(6\pi)}
+{\mathcal{M}}_\text{CT,1PI}^{(6\pi)}\\
&+6(Z^{1/2}-1){\mathcal{M}}_\LO^{(a)(2)}
+{\mathcal{M}}_\LO^{(a)(4)}
\,.
\end{split}
\label{eq:6pi_A4}
\end{equation}
The discussion of the 1PR part is a bit more involved.
The double-pole part ${\mathcal{M}}^\text{2-pole}$ stemming from the contributions represented by diagrams depicted in \cref{fig:6piNLO:2prop,fig:6piNLO:A-2prop} cancels with the piece due to the NLO propagator mass renormalization in the LO pole contribution ${\mathcal{M}}_\LO^{(b)(2)}$ from \cref{fig:6piLO:prop}, denoted as ${\mathcal{M}}_\LO^{(b)(4)}\big|_\text{prop}$, and consequently
\begin{equation}
\begin{split}
&{\mathcal{M}}^\text{2-pole}
+6(Z^{1/2}-1){\mathcal{M}}_\LO^{(b)(2)}
+{\mathcal{M}}_\LO^{(b)(4)}\big|_\text{prop}\\
&=8(Z^{1/2}-1){\mathcal{M}}_\LO^{(b)(2)}\,.
\end{split}
\end{equation}
This, together with the LO contribution itself, ${\mathcal{M}}_\LO^{(b)(2)}$, with vertices promoted to NLO in the same contribution, ${\mathcal{M}}_\LO^{(b)(4)}\big|_\text{vert}$, and contributions stemming from the 1PR topologies shown in \cref{fig:6piNLO:prop,fig:6piNLO:A-prop,fig:6piNLO:B-prop} gives, schematically,
\begin{equation}
\begin{split}
&A_{6\pi}^{(\text{LO+NLO})}\big|_\text{1PR}
={\mathcal{M}}_\text{1-loop,1PR}^{(6\pi)}
+{\mathcal{M}}_\text{CT,1PR}^{(6\pi)}\\
&+8(Z^{1/2}-1){\mathcal{M}}_\LO^{(b)(2)}
+{\mathcal{M}}_\LO^{(b)(4)}\big|_\text{vert}
+{\mathcal{M}}_\LO^{(b)(2)}\,,
\end{split}
\end{equation}
which is the equivalent of two up-to-NLO $\pi\pi$ scatterings [analogous to $A_{4\pi}^{(\LO)}+A_{4\pi}^{(\NLO)}$ as from \cref{eq:A4pi_LO,eq:A4pi_NLO}] connected with the propagator, i.e.\ precisely the structure of \cref{eq:factor}.
Choosing the particular form of $A_{6\pi}^{(\pole)}$ as discussed earlier, the remainder with respect to $A_{6\pi}^{(\text{LO+NLO})}\big|_\text{1PR}$ is deferred to $A_{6\pi}^{(\nonpole)}$.
What we call the nonpole part of the six-meson amplitude is thus the combination of such a remainder and the contributions of the 1PI diagrams from \cref{eq:6pi_A4}.

\subsection{Zero-momentum limit}

In what follows, we choose a symmetric $3\to3$ scattering configuration given by the four-momenta
\begin{equation}
\begin{aligned}
\label{eq:kinematics}
    p_1&=\left(E_p,p,0,0\right),\\
    p_{2,3}&=\left(E_p,-\tfrac{1}{2}p,\pm\tfrac{\sqrt{3}}{2}p,0\right),\\
    p_4&=\left(-E_p,0,0,p\right),\\
    p_{5,6}&=\left(-E_p,\pm\tfrac{\sqrt{3}}{2}p,0,-\tfrac{1}{2}p\right),
\end{aligned}
\end{equation}
with $E_p=\sqrt{p^2+M_\pi^2}$.
These only depend on a single parameter $p$, the modulus of three-momenta of all the mesons.
In this kinematic setting, the zero-momentum limit of the stripped nonpole amplitudes up-to-and-including NLO take a simple group-universal form
\begin{widetext}
    \begin{alignat}{4}\label{eq:p0limit}
        F_\pi^2\lim_{p\to 0}\sampl{6}
            &=&-\frac12\frac{M_\pi^2}{F_\pi^2}
            \:+\:&\frac{M_\pi^4}{F_\pi^4}
            \bigg\{\frac\sgn4(8L-\kappa)
            +(3L-5\kappa)\bigg(\frac{10n+\sgn}{36}-\frac1{\kk n}\bigg)
            +\frac\kappa{2\kk n}
            -4(8L_0^\r-L_5^\r+6L_8^\r)
            \bigg\}\,,\notag\\
        F_\pi^2\lim_{p\to 0}\sampl{2,4}
            &=&&\frac{M_\pi^4}{F_\pi^4}
            \bigg\{
            \frac1\kk(3L+\kappa)
            +2(L-2\kappa)\frac1{\kk^2n^2}
            -16(L_2^\r+L_4^\r+2L_6^\r)
            \bigg\}\,,\notag\\
        F_\pi^2\lim_{p\to 0}\sampl{3,3}
            &=&\:-\frac1{\kk n}\frac{M_\pi^2}{F_\pi^2}
            \:+\:&\frac{M_\pi^4}{F_\pi^4}
            \bigg\{
            \frac3{2\kk}L
        -\frac32(L-\kappa)\bigg(\frac\sgn{\kk n}-\frac4{\kk^2n^2}\bigg)
        -16L_7^\r-\frac{48}{\kk n}(L_5^\r-L_8^\r)
            \bigg\}\,,\notag\\
        F_\pi^2\lim_{p\to 0}\sampl{2,2,2}
            &=&&\frac{M_\pi^4}{F_\pi^4}\bigg\{\frac{4\kappa}{\kk^3n^3}\bigg\}\,.
    \end{alignat}
    % \pagebreak % to avoid a silly little one-line residue after the equation
\end{widetext}
Due to the Adler zero $\lim_{\epsilon\to 0}A(p_1,\ldots,\epsilon p_i,\ldots) = 0$, which holds for any $i$ in the massless case~\cite{Adler:1964um,Adler:1965ga}, the zero-momentum limit is proportional to $M^2_\pi$ also in the general case.

It seems that \cref{eq:p0limit} is valid also for general momentum configurations rather than just \cref{eq:kinematics}; this is explained in the next section.
However, it is specifically the zero-momentum limit of $\ampl_R(p_1,\ldots,p_6)$ where particles $1,2,3$ are in the initial state and $3,4,5$ in the final state.
Different assignments of initial- and final-state particles will yield different zero-momentum limits.
After accounting for $\Z_R^\rev$ symmetry, time-reversal symmetry, and the freedom to exchange particles within the initial and final states (which changes the stripped amplitude but not its zero-momentum limit%
\footnote{
    This is not necessarily true if the limit depends nontrivially on how it is approached.
    This seems not to be the case for this amplitude, although the analytic structure of $C$ could hide some subtleties.}%
), there are 10 distinct limits, produced by the following:
\begin{equation}\label{eq:different-limits}
    \begin{aligned}
        &\left.\begin{gathered}
             \lim_{p\to 0}\ampl_R(p_1,p_2,p_3,p_4,p_5,p_6)\,,\\
             \lim_{p\to 0}\ampl_R(p_1,p_4,p_2,p_5,p_3,p_6)\,,
        \end{gathered}
        \quad\right\}\;\text{for all $R$}\,,\\
        &\lim_{p\to 0}\sampl{6}(p_1,p_2,p_4,p_3,p_5,p_6)\,,\\
        &\lim_{p\to 0}\sampl{2,4}(p_1,p_4,p_2,p_3,p_5,p_6)\,.
    \end{aligned}
\end{equation}
Note that the first line reproduces \cref{eq:p0limit}; in the interest of space, we do not reproduce the other cases.
Also note that this is for $3\to3$ scattering, and that different limits will be obtained for $2\to4$ scattering. 

\section{Numerical results}
\label{sec:numerics}

We only present a few numerical results here since the full analysis of the finite volume and the subtraction of the two-body rescatterings is very nontrivial; see \rrcite{Romero-Lopez:2020rdq,Blanton:2020jnm} and references therein.
The numerical inputs we use are
\begin{equation}
    \begin{aligned}
        M_\pi&=0.139570\,\text{GeV}\,,& \mu&=0.77\,\text{GeV}\,,\\
        F_\pi&=0.0927\,\text{GeV}\,,&   n&=3\,.
    \end{aligned}
\end{equation}
For LECs, we use the $p^4$ fit from Table~1 of \rcite{Bijnens:2014lea}:
\begin{equation}
    \begin{aligned}
        L_1^\r&=1.0\times10^{-3}\,,& L_5^\r&=1.2\times10^{-3}\,,\\
        L_2^\r&=1.6\times10^{-3}\,,& L_6^\r&=0\,,\\
        L_3^\r&=-3.8\times10^{-3}\,,& L_7^\r&=-0.3\times10^{-3}\,,\\
        L_4^\r&=0\,,& L_8^\r&=0.5\times10^{-3}\,.
    \end{aligned}
\end{equation}
For $n=3$, we use $L_0^\r=0$.
Throughout this section, we use the kinematic setting of \cref{eq:kinematics}.

\begin{figure*}[t]
    \begin{subfigure}[t]{0.48\linewidth}
        \includegraphics[width=\linewidth]{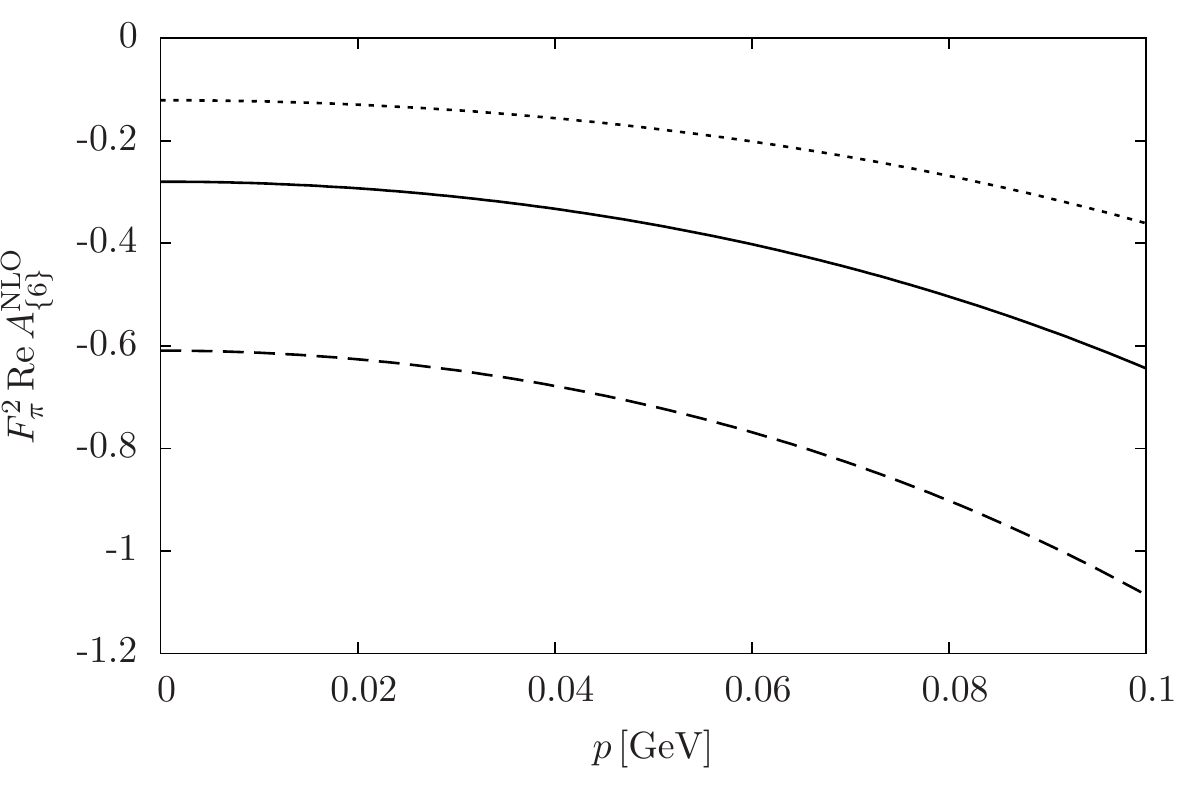}
    \end{subfigure}
    \hfill
    \begin{subfigure}[t]{0.48\linewidth}
        \includegraphics[width=\linewidth]{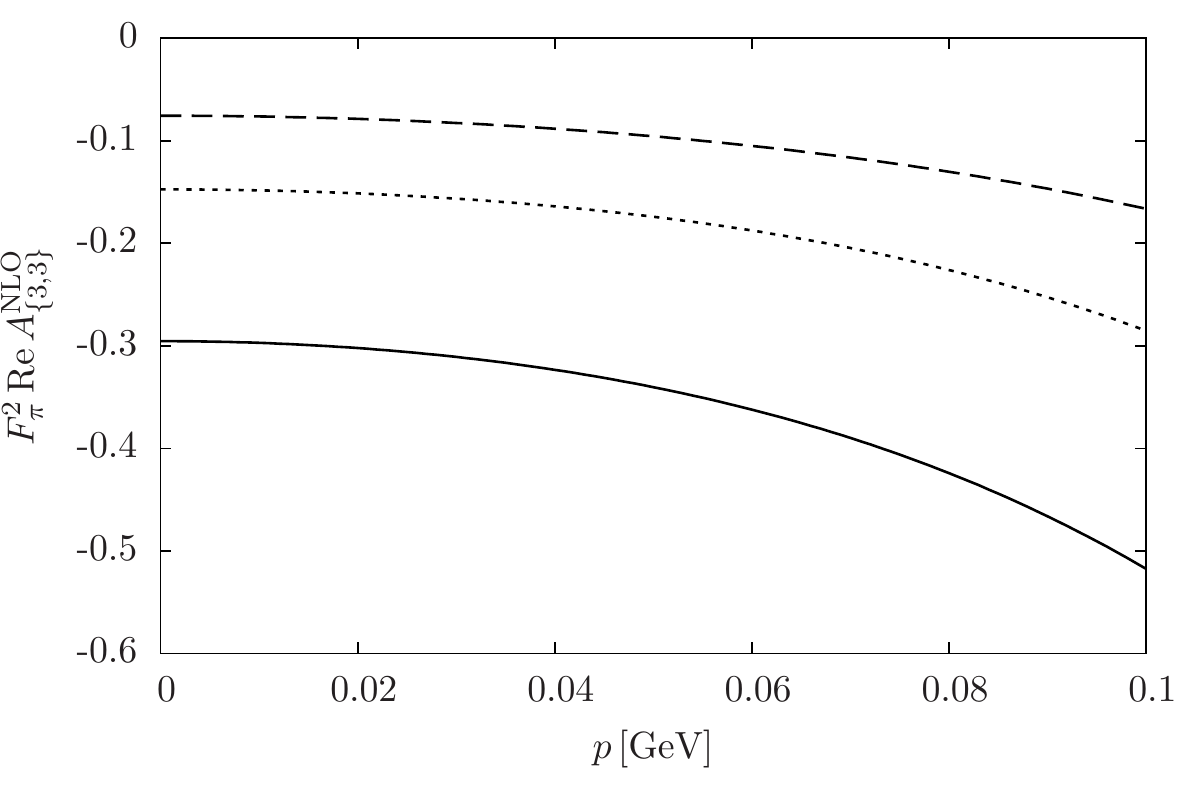}
    \end{subfigure}

    \begin{subfigure}[t]{0.48\linewidth}
        \includegraphics[width=\linewidth]{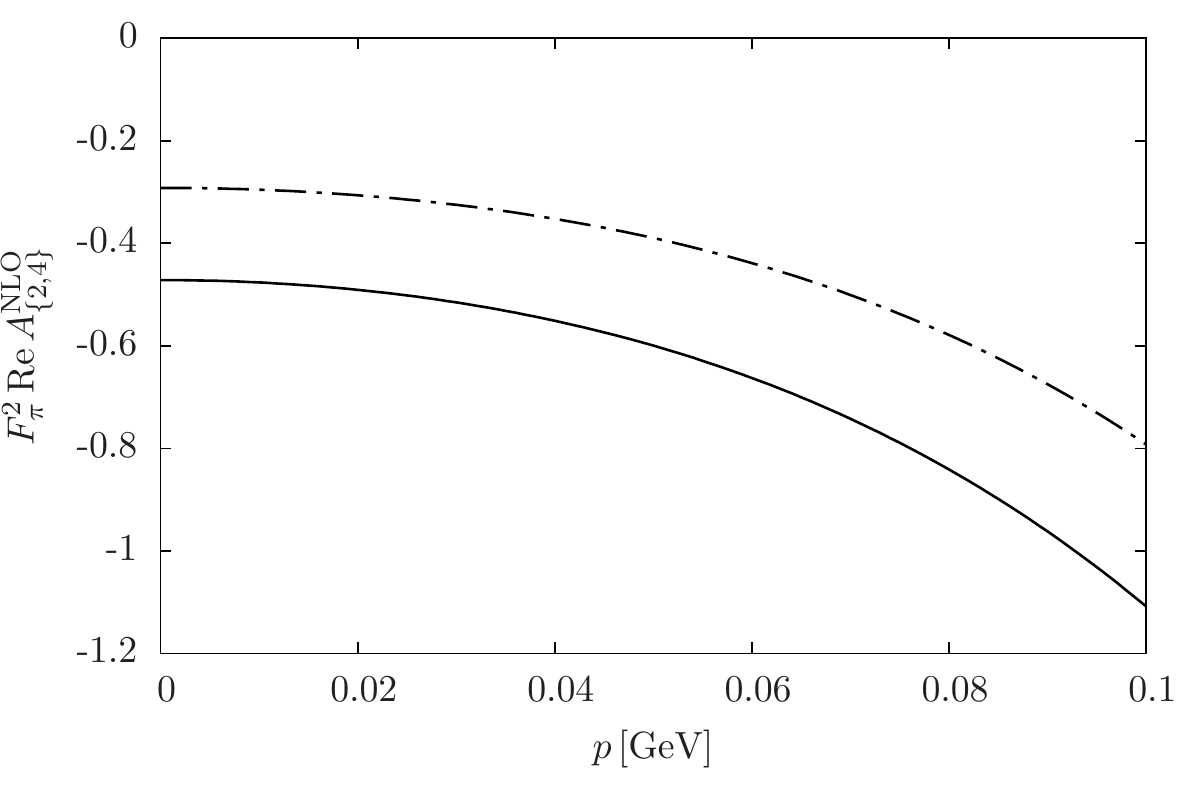}
    \end{subfigure}
    \hfill
    \begin{subfigure}[t]{0.48\linewidth}
        \includegraphics[width=\linewidth]{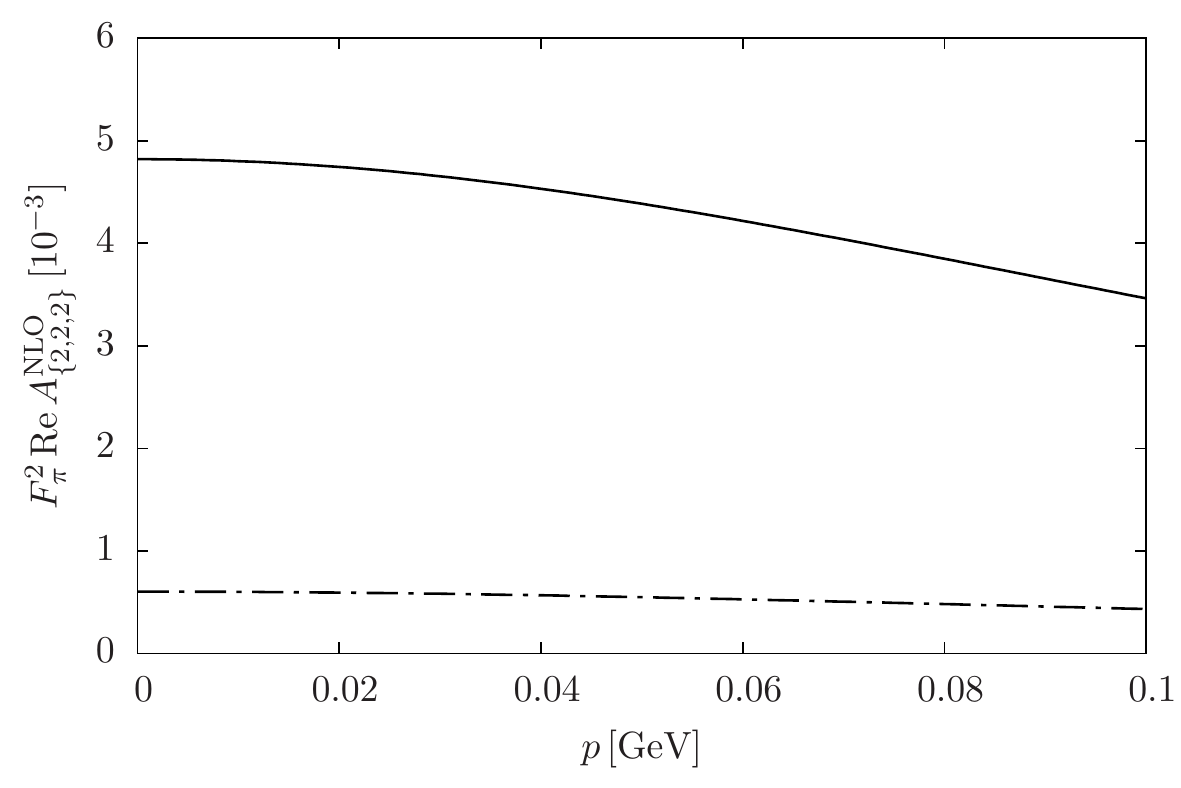}
    \end{subfigure}
    \caption{
        Stripped amplitudes at NLO in the kinematic setting of \cref{eq:kinematics}.
        The solid line stands for $\SU$, the dashed line for $\SO$ and the dotted line for $\Sp$.
        The dash-dotted line used in the two bottom panels represents the cases in which $\SO$ and $\Sp$ coincide.
        Note the extra factor of $10^{-3}$ for the $R=\{2,2,2\}$ stripped amplitude.}
        \label{fig:plots}
\end{figure*}

The resulting plots are shown in \cref{fig:plots}.
Interestingly, $\sgampl{3,3}{\NLO,\sgnp}$ vanishes in this kinematic setting and hence does not contribute to the top right panel of \cref{fig:plots}.

\begin{table}[tb]
    \setlength{\tabcolsep}{3pt}
    \renewcommand{\arraystretch}{1.7}
    \centering
    \begin{tabular}{c|c|c}
        \toprule
        & \multicolumn{1}{c|}{Pole part} &   \multicolumn{1}{c}{Non-pole part}\\
        \midrule
            $\hsgampl{6}{\LO,1}$        & $86.4$                            & $-1.68$ \\
            $\hsgampl{3,3}{\LO,1}$      & $5.06\tfrac1n$                    & $-4.89\tfrac1n$ \\
        \midrule
            $\hsgampl{6}{\NLO,1}$       & $-36.2 - 10.2\tfrac1n$            & $-0.0119 + 0.477\tfrac1n$\\
            $\hsgampl{6}{\NLO,\sgn}$    & $0.248$                           & $-0.363$ \\
            $\hsgampl{6}{\NLO,\kk}$     & $12.4\,n$                         & $-0.264\,n$\\
            $\hsgampl{2,4}{\NLO,1}$     & $20.5+4.84\tfrac1{n^2}$           & $-0.492 - 0.293\tfrac1{n^2}$\\
            $\hsgampl{2,4}{\NLO,\kk}$   & $24.4$                            & $-0.584$\\
            $\hsgampl{3,3}{\NLO,1}$     & $1.75\tfrac1n+0.624\tfrac1{n^2}$  & $0.0247 - 0.606\tfrac1n - 0.741\tfrac{1}{n^2}$\\
            $\hsgampl{3,3}{\NLO,\sgn}$  & $-0.328\tfrac1n$                  & $0.357\tfrac1n$\\
            $\hsgampl{3,3}{\NLO,\sgnp}$ & ---                               & $0$\\
            $\hsgampl{3,3}{\NLO,\kk}$   & $-0.501$                          & $-0.258$\\
            $\hsgampl{2,2,2}{\NLO,1}$   & ---                               & $0.0935\tfrac1{n^3}$\\
        \bottomrule
    \end{tabular}
    \caption{
        The different contributions to the six-meson amplitude, evaluated at $p=0.1$\,GeV as described in \cref{eq:hat}, using the momentum configuration~\eqref{eq:kinematics}.
        Note that we only quote the real part, and that we multiply by a suitable power of $F_\pi$ to make the result dimensionless.
        We omit subamplitudes that are identically zero.}
    \label{tab:numerics}
\end{table}

We use the following shorthand notation for the value at $p=0.1$\,GeV:
\begin{equation}\label{eq:hat}
    \ampl(p)\;\longrightarrow\;\hampl \equiv F_\pi^2\times\operatorname{Re}\ampl(p=0.1\,\text{GeV})\,.
\end{equation}
Note that $\hampl$ is dimensionless.
Using this notation, values for general $n$ are given in \cref{tab:numerics}.
As \cref{fig:plots} shows, the relative sizes of these values are representative across a broader energy range.

The pole part is clearly the dominant contribution, but in a sense the nonpole part is the interesting one, since it is not directly related to the previously known four-meson amplitude.
The NLO part is smaller than the LO part, but not by much; perturbative convergence is understandably poor, with breakdown expected at a scale of $4\pi F_\pi/\sqrt{n}$~\cite{Chivukula:1992gi}, i.e.\ ${\approx}\,5M_\pi$ at $n=3$.

Among the NLO nonpole subamplitudes, neither is clearly dominant: $\sampl{6}$, $\sampl{2,4}$ and $\sampl{3,3}$ are comparable in magnitude, as are $\gampl{1}$, $\gampl{\sgn}$ and $\gampl{\kk}$.
Among the NLO pole subamplitudes, $\sgampl{6}{1},\sgampl{6}{\kk}$ and $\sgampl{2,4}{1}$ dominate.
Thus, the results of the three QCD-like theories are of similar magnitude.
In the large-$n$ limit, we expect $\sampl{6}$ to dominate due to the positive power of $n$ in $\sgampl{6}{\kk}$; following \cref{eq:group-univ}, this also implies that the stripped amplitudes of the three theories will become equal in this limit.

Besides the kinematic configuration~\eqref{eq:kinematics}, we have numerically evaluated the amplitude for a random sample of $3\to3$ scattering events generated with the RAMBO algorithm~\cite{Kleiss:1985gy}.
These samples confirm that $\sgampl{3,3}{\NLO,\sgnp}$ is not generally zero.
We obtained zero-momentum limits by uniformly scaling the random three-momenta by a factor $\epsilon\to0$ while keeping the particles on-shell.%
\footnote{
    This introduces an energy-conservation-violating term of order $\epsilon^2$.
    Violating either on-shellness or conservation of energy (or momentum) is inevitable when taking such a zero-momentum limit.
    It would be simpler to scale the four-momenta, but that would give the massless zero-momentum limit, which is zero and therefore not very interesting.}
This consistently resulted in the same numerical values as \cref{eq:p0limit}, allowing us to conclude that \cref{eq:p0limit} is the general uniform zero-momentum limit of $A_R(p_1,\ldots,p_6)$ in $3\to3$ scattering, rather than a special case for the configuration~\eqref{eq:kinematics}.
The same is true for the other limits described in \cref{eq:different-limits}.

\section{Conclusions}
\label{sec:conclusions}

In this paper, we calculated the meson mass, decay constant, four-meson and six-meson amplitudes to NLO in the QCD-like theories, i.e.\ besides the QCD-related case $\SU$, we also consider the $\SO$ and $\Sp$ symmetry-breaking patterns.
In \cref{sec:theory}, we introduced the relevant NLO Lagrangian in analogy with three\mbox{(-quark)}-flavor ChPT~\cite{Gasser:1983yg,Gasser:1984gg}.
The mass, decay constant and the four-meson amplitude agree for $n=3$ with \rrcite{Gasser:1983yg,Gasser:1984gg,Gasser:1986vb,Splittorff:2001fy,Bijnens:1995yn,Bijnens:1997vq,Bijnens:2009qm} and with the general-$n$ results of \rrcite{Dobado:1994fd,Bijnens:2009zi,Bijnens:2010xg,Bijnens:2011fm,Chivukula:1992gi}.

Our main result is the six-meson amplitude, which can be written in terms of four independent flavor-stripped amplitudes [for detailed structure, see \cref{eq:DEFG}], as compared to a single amplitude in the $\OO(N+1)/\OO(N)$ case studied in \rcite{Bijnens:2021hpq}.
We split the whole amplitude into pole and nonpole parts; see \cref{eq:pole-nonpole}.
The pole part is given in \cref{eq:factor}, where we chose to employ the off-shell four-meson amplitude in the form of \cref{eq:A4pi_LO,eq:4meson} generalizing (beyond $n=3$) the amplitude given in \rrcite{Bijnens:1995yn,Bijnens:1997vq} and exactly matching that in \rcite{Bijnens:2011fm}.

The expression for the nonpole part is rather lengthy.
We thus further divide the four flavor-stripped amplitudes into group-universal subamplitudes in order to account for all the three QCD-like theories in a concise way.
By employing symmetries through the deorbiting procedure described in \cref{app:deorbit}, we obtain the resulting 10 non-vanishing subamplitudes presented in \cref{app:6pi-res}.
The nontrivial choice of a redundant but highly symmetric basis of tensor triangle loop integrals (for details, see \cref{app:integrals}) and of kinematic invariants (\cref{app:deorbit}) allows for a fairly compact expression.
While the result is still too lengthy and complicated to be grasped fully, the division into subamplitudes along with further analysis in \cref{app:groups} allows many of its features to be understood.

In the kinematic setting of \cref{eq:kinematics}, we present the analytical results for the zero-momentum limit in \cref{eq:p0limit}.
Some numerical results for this particular momentum configuration are presented in \cref{sec:numerics}.

In the process of our calculations, we devised a systematic procedure (deorbiting) for simplifying amplitudes beyond what is possible with stripping alone.
Previous work, e.g.\ \rrcite{Bijnens:2011fm,Kampf:2013vha,Bijnens:2021hpq}, manually structure their results in similar ways, but this quickly becomes difficult with larger numbers of kinematic variables and more complicated symmetries.
These issues are, at least partly, resolved by our simplification scheme, which should be applicable also beyond the present scope.

We see limited interest in computing the NNLO counterpart of this result.
Several LECs (terms 49--63 in $\mathcal L^{(6)}$~\cite{Bijnens:1999sh}) that do not appear in lower-multiplicity amplitudes enter here and are so far undetermined.
All relevant two-loop integrals are known (see e.g.\ \rcite{Bijnens:2011fm}) except for the five-propagator sunset topology, which we expect to be very difficult.
There is also the matter of expressing the two-loop integrals in a symmetry-compliant way like in \cref{app:integrals}.

We believe that our techniques would make the NLO eight-meson amplitude accessible, but such a calculation is currently not motivated by lattice developments. 
Besides the larger number of diagrams and longer expressions, the main technical hurdles would be extending \cref{app:integrals} to a similar treatment of box integrals, and extending \cref{app:deorbit} to $\Z_{\{8\}}^\rev,\Z_{\{2,6\}}^\rev,\Z_{\{2,3,3\}}^\rev$, etc. 

Work is in progress to combine our results with the methods for extracting three-body scattering from finite volume in lattice QCD.
We expect that our results may also be of interest for the amplitude community.

\begin{acknowledgments}
The authors thank R.~Frederix and A.~Lifson for suggestions about random momentum sampling.
This work is supported in part by the Swedish Research Council grants contracts no.~2016-05996 and no.~\mbox{2019-03779}.
\end{acknowledgments}

\appendix

\renewcommand{\thesubsection}{\Alph{section}--\arabic{subsection}}

\section{Conventions for the loop integrals}
\label{app:integrals}

Throughout the paper, we treat the momenta ($p_1,\,\dots,\,p_6$) as incoming, and we introduce the following independent combinations of momenta:
\begin{equation}
    \begin{alignedat}{6}
        q_1&=p_1+p_2\,,\quad&  q_2&=p_3+p_4\,,\quad&  q_3&=p_5+p_6\,,\\
        r_1&=p_1-p_2\,,\quad&  r_2&=p_3-p_4\,,\quad&  r_3&=p_5-p_6\,.
    \end{alignedat}
\end{equation}
Note that we use the same notation and conventions as in \rcite{Bijnens:2021hpq}.
In particular, the integrals are defined in Appendix A therein;
here, we restate them along with some clarifications.

The functions we use to represent our results are very closely related to the standard Passarino--Veltman one-loop integrals $A_0$, $B_0$ and $C_0$.
To fix our notation, let us present explicitly the simpler integrals with one and two propagators.
In what follows, we use the compact notation for the Feynman denominators with loop momentum $l$,
\begin{equation}
    D(q_i) \equiv(l-q_i)^2-M^2\,,
\end{equation}
while setting $D_0\equiv D(0)=l^2-M^2$.
Having in mind \cref{eq:epstilde},
\begin{equation}
    \frac1{\tilde\epsilon} \equiv\frac{1}{\epsilon}-\gamma_\text{E}+\log4\pi-\log\mu^2+1\,,
\end{equation}
and that we, like in \cref{sec:theory}, set
\begin{align}
  \kappa    &=\frac{1}{16\pi^2}\,,&
  L         &=\kappa\log\frac{M^2}{\mu^2}\,,
\end{align}
the integrals read
\begin{align}
    \label{eq:AB}
    A
    &\equiv\kappa A_0(M^2)
    =\frac1i\int\frac{\text{d}^dl}{(2\pi)^d}\frac{1}{D_0}\notag\\
    &=M^2\kappa\frac1{\tilde\epsilon}-M^2 L\,,\\
    % \begin{split}
    B(q^2)
    &\equiv\kappa B_0(q^2,M^2,M^2)
    =\frac{1}{i}\int\frac{\text{d}^dl}{(2\pi)^d}\frac{1}{D_0D(q)}\notag\\
    &=\kappa\frac1{\tilde\epsilon}-\kappa-L+\bar J(q^2)\,.
    % \end{split}
\end{align}
We employ the standard definition for $\bar J(q^2)$:
\begin{equation}
    \bar J(q^2)
    \equiv\kappa\bigg(2+\beta\log\frac{\beta-1}{\beta+1}\bigg)\,,
\end{equation}
with $\beta\equiv\beta(q^2)=\sqrt{1-\frac{4M^2}{q^2}}$.
The terms $L$ and $\bar J(q^2)$ we use to express our results thus absorb the factors of $\frac1{16\pi^2}$.

Let us emphasize that it is the tensor triangle one-loop integrals of higher ranks which generate lengthy expressions upon reduction to the scalar ones.
It therefore turned out to be more convenient to use a specific basis for the tensor integrals with particular symmetry properties.
Regarding the rank-3 integrals, the combination
\begin{widetext}
    \begin{equation}
        \label{eq:C3}
        C_3(p_1,p_2,\dots,p_6)
        =\frac{1}{3}\frac{1}{i}\int\frac{\text{d}^dl}{(2\pi)^d}
          \frac{l\cdot r_1 \,l\cdot r_2\, l\cdot r_3}{D_0}\\
        \bigg[\frac{1}{D(q_1)D(-q_2)}+\frac{1}{D(q_2)D(-q_3)}+\frac{1}{D(q_3)D(-q_1)}\bigg]
    \end{equation}
\end{widetext}
has more symmetries than the first term only and is UV finite.
It is antisymmetric under the interchange of the momenta inside each pair [pairs being here $(p_1,p_2)$, $(p_3,p_4)$ and $(p_5,p_6)$] and
antisymmetric under the interchange of two pairs.
The rank-2 integral can be defined as
\begin{equation}
  C_{21}(p_1,p_2,\dots,p_6)
  =\frac{1}{i}\int\frac{\text{d}^dl}{(2\pi)^d}
  \frac{l\cdot r_1 \,l\cdot r_2}
  {D_0D(q_1)D(-q_2)}\,.
\end{equation}
It is antisymmetric under the interchange $p_1\leftrightarrow p_2$ and symmetric under $(p_1,p_2)\leftrightarrow(p_3,p_4)$ and $p_5\leftrightarrow p_6$.
The integral with one product $l\cdot r_i$ in the numerator can be defined as
\begin{equation}
  C_{11}(p_1,p_2,\dots,p_6)
  =\frac{1}{i}\int\frac{\text{d}^dl}{(2\pi)^d}
  \frac{l\cdot r_3}{D_0D(q_1)D(-q_2)}\,.
\end{equation}
It is antisymmetric under the interchange $p_5\leftrightarrow p_6$, $(p_1,p_2)\leftrightarrow(p_3,p_4)$ and is symmetric under $p_1\leftrightarrow p_2$ and $ p_3\leftrightarrow p_4$.
Owing to the symmetries, other integrals of ranks 2 and 1 can be expressed in terms of $C_{21}$ and $C_{11}$ and integrals with lower ranks, respectively, so we only need those to write out our final result.
Finally, we define
\begin{equation}
  C(p_1,p_2,\dots,p_6)
  =\frac{1}{i}\int\frac{\text{d}^dl}{(2\pi)^d}\frac{1}{D_0D(q_1)D(-q_2)}\,,
\end{equation}
which is symmetric under $p_1\leftrightarrow p_2$ and under all pair interchanges.
It is related to $C_0$ as
\begin{equation}
    C(p_1,p_2,\dots,p_6)
    =\kappa C_0(q_1^2,q_2^2,q_3^2,M^2,M^2,M^2)\,,
\end{equation}
in which case the mentioned symmetries are seen trivially due to equal masses.

We express the amplitude in terms of $C_3$, $C_{21}$, $C_{11}$ and $C$.
As already mentioned, the former three can be expressed in terms of $C$, but the expressions are cumbersome and lead to a very long expression for the amplitude.
We have therefore kept all these four, among which only $C_{21}$ contains a UV-infinite part:
\begin{equation}
  C_{21}(p_1,p_2,\ldots,p_6)
  =\kappa\,\frac{r_1\cdot r_2}{4}
  \frac1{\tilde\epsilon}+\overline C_{21}(p_1,p_2,\ldots,p_6)\,.
\end{equation}  

\section{General parametrization}
\label{app:params}

In this appendix, we show how to parametrize a special unitary matrix $u$ in full generality.
Special cases of what follows give parametrizations such as the four used in \rcite{Bijnens:2013yca}.

A special unitary matrix $\uu$ can always be written as an exponential of a Hermitian traceless matrix $\phi$:
\begin{equation}
    \uu = \exp(i\phi)\,,\quad\phi^\dagger=\phi\,,\quad\tr{\phi}=0\,.
\end{equation}
The obvious way to write a general parametrization is
\begin{equation}
    \uu \longrightarrow \sum_{m=0}^\infty b_m \phi^m\,,
\end{equation}
with unitarity conditions relating the $b_m$.
As proven in \rcite{Cronin:1967jq}, the only generally valid solution where $b_m$ are c-numbers is $\uu=\exp(i\phi)$, although sufficiently low order, other such parametrizations are valid and useful; see e.g.\ \rrcite{Ellis:1970nt,Kampf:2013vha}.
Generally, however, $b_m$ are functions of traces of powers of $\phi$, as seen in \rcite{Bijnens:2013yca}; this complicates the unitarity conditions.

Here, we take the alternative approach of redefining $\phi$ to $\phi'(\phi)$ with ${\phi'}^\dagger={\phi'}$ and $\tr{{\phi'}}=0$, and keeping $\uu=\exp(i\phi')$.
Under the unbroken (vector) part of the chiral transformation, $\phi\to g_\text{V}\phi g_\text{V}^\dagger$, and we want ${\phi'}\to g_\text{V}{\phi'} g_\text{V}^\dagger$.
The redefined ${\phi'}$ is thus a series in $\phi$ and traces of powers of $\phi$:
\begin{align}\label{eq:reparam}
    {\phi'}
    =\phi+\hspace{-.7cm}\sum_{\substack{
    m\ge2,\,j\ge 0,\,\\
    i_0+\cdots+i_j=m,\,\\
    i_0\ge0;\:i_1\ge i_2\ge\,\cdots\,i_j\ge2
    }}\hspace{-.7cm}
    a_{i_0\ldots i_j}\phi^{i_0}\tr{\phi^{i_1}}\cdots\tr{\phi^{i_j}}\,.
\end{align}
Further restrictions follow from using intrinsic parity, i.e.\ employing ${\phi'}\to -{\phi'}$ if $\phi\to-\phi$, thus allowing only for odd values of $m$, and applying ${\phi'}^\dagger={\phi'}$, which requires the $a_{i_0\ldots i_j}$ to be real.
The condition \mbox{$\tr{{\phi'}}=0$} determines all the $a_{0 i_1\ldots i_j}$ (those with $i_0=0$) to be $a_{0 i_1\ldots i_j}=-a_{i_1\ldots i_j}/\kk n$.
Hence, all terms relevant for the six-meson amplitude discussed in this work, introducing 18 extra unconstrained parameters,%
\footnote{
    In general, the number of free parameters at order $\phi^m$ is equal to the number of ways to partition $m-1$ into \mbox{positive} integers, i.e.\ Online Encyclopedia of Integer Sequences (OEIS) sequence A058696 starting with $2,5,11,22,43,77,135$.
    Let us briefly sketch the proof of this.
    Let $i_0\geq 1$ since $a_{0i_1\ldots i_j}=-a_{i_1\ldots i_j}/\kk n$.
    Then, rewrite each term in \cref{eq:reparam} as
    \begin{equation*}
        a_{i_0i_1,\ldots i_j} \phi \tr{\phi^{i_1}}\tr{\phi^{i_2}}\cdots\tr{\phi^{i_j}} \underbrace{\phi\phi\cdots\phi}_{\text{$i_0-1$ $\phi$'s}}
    \end{equation*}
    In total, there are $m$ factors of $\phi$, of which all but the first can be arbitrarily partitioned like $m-1\to i_1,\ldots,i_j,1,\ldots,1$.
    Since $\tr{\phi}=0$ and $i_1\geq i_2\geq\,\cdots\,i_j\geq 2$, we can unambiguously associate the 1's with $\phi$'s outside traces and the other elements of the partition with $i_1,\ldots i_j$.
    This demonstrates the one-to-one correspondence between independent parameters $a_{i_0i_1,\ldots i_j}$ and partitions of $m-1$.}
are in group-universal form
\begin{widetext}
    \begin{align}
        {\phi'}
            =\phi
            &+a_3\Big(\phi^3-\tfrac{1}{\kk n}\tr{\phi^3}\Big)
            +a_{12}\,\phi\tr{\phi^2}\notag\\
            &+a_5\Big(\phi^5-\tfrac{1}{\kk n}\tr{\phi^5}\Big)
            +a_{32}\Big(\phi^3\tr{\phi^2}-\tfrac{1}{\kk n}\tr{\phi^3}\tr{\phi^2}\Big)
            +a_{23}\Big(\phi^2\tr{\phi^3}-\tfrac{1}{\kk n}\tr{\phi^3}\tr{\phi^2}\Big)
            +a_{14}\,\phi\tr{\phi^4}
            +a_{122}\,\phi\tr{\phi^2}^2\notag\\
            &+a_7\Big(\phi^7-\tfrac{1}{\kk n}\tr{\phi^7}\Big)
            +a_{52}\Big(\phi^5\tr{\phi^2}-\tfrac{1}{\kk n}\tr{\phi^5}\tr{\phi^2}\Big)
            +a_{43}\Big(\phi^4\tr{\phi^3}-\tfrac{1}{\kk n}\tr{\phi^4}\tr{\phi^3}\Big)\notag\\
            &+a_{34}\Big(\phi^3\tr{\phi^4}-\tfrac{1}{\kk n}\tr{\phi^4}\tr{\phi^3}\Big)
            +a_{322}\Big(\phi^3\tr{\phi^2}^2-\tfrac{1}{\kk n}\tr{\phi^3}\tr{\phi^2}^2\Big)
            +a_{25}\Big(\phi^2\tr{\phi^5}-\tfrac{1}{\kk n}\tr{\phi^5}\tr{\phi^2}\Big)\notag\\
            &+a_{232}\Big(\phi^2\tr{\phi^3}\tr{\phi^2}-\tfrac{1}{\kk n}\tr{\phi^3}\tr{\phi^2}^2\Big)
            +a_{16}\,\phi\tr{\phi^6}
            +a_{142}\,\phi\tr{\phi^4}\tr{\phi^2}
            +a_{133}\,\phi\tr{\phi^3}^2
            +a_{1222}\,\phi\tr{\phi^2}^3\notag\\
            &+\mathcal{O}\big(\phi^9\big)\,.
    \end{align}
\end{widetext}
The presence of traces in all terms except the first confirms and generalizes the conclusions of \rcite{Cronin:1967jq}.

Taking this general form to define $u$ via \cref{eq:u_exp} with $\phi \equiv \phi^at^a/\sqrt2 F$ and plugging it into the Lagrangian~\eqref{eq:L} adds an extra cross-check of one's calculations, since the physical amplitude cannot depend on $a_{i_0\ldots i_j}$.

\section{Deorbiting and closed bases of kinematic invariants}
\label{app:deorbit}

In this appendix, we briefly describe the method used for the final simplification step of reducing $\gampl{i}_R$ to $\dgampl{i}_R$ with the property~\eqref{eq:deorbit}.
This is a development of an \emph{ad hoc} technique used in \rcite{Bijnens:2019eze}.
Recall that $\dgampl{i}_R$ is not unique and that our aim is to make its expression as short as possible.

Consider some class of objects $x$ and a group $\mathcal G$ (in our case, $x$ are products of one-loop integral functions and kinematic invariants, and $\mathcal G = \Z^\rev_R$).
In standard nomenclature, the set of objects obtained by acting with  $\mathcal G$ on $x$ is called the \emph{orbit} of $x$, denoted $\mathcal G\cdot x$; formally,
\begin{equation}
    \mathcal G\cdot x = \left\{g\cdot x\,\middle|\, g\in\mathcal G\right\}.
\end{equation}
Consider then an expression $X$ composed of a sum of objects $x$.
Reducing it to $\tilde X$ such that $X=\sum_{g\in\mathcal G}g\cdot\tilde X$ (where $g\cdot\tilde X$ indicates acting with $g$ on each term in $\tilde X$) is done using the following algorithm:
\begin{enumerate}
    \item Start with $\tilde X=0$.
    \item Select the first term $x$ in $X$, under some arbitrary but consistent ordering of the terms.%
    \footnote{
        In practice, we use the internal ordering of FORM, with some modifications.}
    \item Compute the orbit $\mathcal G\cdot x$ and the symmetry factor $S = |\mathcal G|/|\mathcal G\cdot x|$ (this is always an integer).
    \item Add $x$ to $\tilde X$, and subtract $\frac1S\sum_{g\in\mathcal G}g\cdot x$ from $X$.
        (Now, no element of $\mathcal G\cdot x$ appears in $X$.)
    \item Repeat from step 2 until $X=0$.
\end{enumerate}
The symmetry factor compensates for how each element of $\mathcal G\cdot x$ appears $S$ times in the sum $\sum_{g\in\mathcal G}g\cdot x$.

Optimally, each object $x$ that appears in any orbit should be a single term, not a sum of other objects.
We will call this property being \emph{closed} under $\mathcal G$.
Without this property, the algorithm may yield poor results or not terminate at all.
However, if the class of objects is closed under $\mathcal G$, it is easy to see that no orbits overlap and that the algorithm results in an $\tilde X$ that is the shortest possible subexpression of $X$, granted that there are no additional symmetries that are not taken into account.

In the context of our amplitudes, we therefore need to carefully choose our basis of kinematic invariants.
We define them in terms of \emph{generalized Mandelstam variables}
\begin{equation}
    \mandel{ij\ldots} \equiv (p_i+p_j+\ldots)^2\,,
\end{equation}
which relate to the usual ones as
\begin{equation}
    s = \mandel{12},\quad t = \mandel{13},\quad u = \mandel{23}.
\end{equation}
Thus, the pairs $s,u$ and $t,u$ form closed bases under $\Z^\rev_R$ for $R=\{4\}$ and $\{2,2\}$, respectively.
The former generalizes straightforwardly to all $R=\{2\N\}$, including the $R=\{6\}$ basis $s_1,\ldots,s_9$ with%
\footnote{
    This basis is valid for 5 or more space-time dimensions; with 4, the correct number of kinematic degrees of freedom is 8, not 9.
    However, the 9th variable is related to the other 8 through the nonlinear Gram determinant relation, so for the sake of simplicity we ignore this and use 9-element bases.
    
    For a $\N$-particle process in $d$ dimensions, a similar basis of generalized Mandelstam variables will have $\N(\N-3)/2$ elements, as is easily found by counting products $p_i\cdot p_j$ and accounting for $p_i^2=M_\pi^2$ and $\sum_i p^\mu_i = 0$.
    This is redundant when $k<d-1$; then, with only $d-1$ independent components in each $p_i$, the number of kinematic degrees of freedom after accounting for $\sum_i p^\mu_i = 0$ and subtracting the dimension of the Lorentz group gives $(d-1)\N - d(d+1)/2$, i.e.\ $3\N-10$ when $d=4$.
    This na\"ive counting does not apply for large $d$, since the momentum vectors live in a $(k-1)$-dimensional subspace; thus, there are $\N(\N-3)/2$ degrees of freedom when $k\geq d-1$.}
\begin{equation}
    \begin{gathered}
            s_1 = \mandel{12},\; s_2 = \mandel{23},\;\ldots,\; s_5 = \mandel{56},\; s_6 = \mandel{61},   \\
            s_7 = \mandel{123},\quad s_8 = \mandel{234},\quad s_9 = \mandel{345}.   
    \end{gathered}
\end{equation}
Finding closed bases is much harder in the other cases.
For $R=\{2,4\}$ we use the basis%
\footnote{
    This is a new basis; the one in \rcite{Bijnens:2019eze} is not closed under trace-reversal.
    It was obtained using similar methods to the $\{3,3\}$ and $\{2,2,2\}$ bases derived in that paper (note that $\Z^\rev_{\{2,4\}}$, unlike $\Z_{\{2,4\}}$, is non-Abelian).}
\begin{equation}
    \begin{gathered}
        \begin{alignedat}{8}
            t_1 &= \mandel{123},\quad& t_2 &= \mandel{124},\;& t_3 &= \mandel{125},\;& t_4 &= \mandel{126},   \\
            t_5 &= \mandel{234},\quad& t_6 &= \mandel{245},\quad& t_7 &= \mandel{256},\quad& t_8 &= \mandel{263},   
        \end{alignedat}\\
        t_9 = \mandel{135} - \mandel{146},
    \end{gathered}
\end{equation}
and for $R=\{3,3\}$%
\footnote{
    This is quite different from the one used in \rcite{Bijnens:2019eze} and is much simpler --- it is just one of the nonets formed under $\Z_{\{3,3\}}$.
    The reason for it being overlooked can be traced back to \rcite{masterthesis}, where an effort was made to include the element $\mandel{123}$ in the basis.}
\begin{equation}
    \begin{alignedat}{6}
        u_1 &= \mandel{14},\qquad    & u_4 &= \mandel{25},\qquad  & u_7 &= \mandel{36},\\
        u_2 &= \mandel{24},\qquad    & u_5 &= \mandel{35},\qquad  & u_8 &= \mandel{16},\\
        u_3 &= \mandel{34},\qquad    & u_6 &= \mandel{15},\qquad  & u_9 &= \mandel{26}.
    \end{alignedat}
\end{equation}
No basis is needed for $R=\{2,2,2\}$ here due to the simplicity of $\sampl{2,2,2}$.
\RRcite{Bijnens:2019eze,Bijnens:2021hpq} provide two different $R=\{2,2,2\}$ bases.

There is no need to apply similar considerations to the loop integrals $\bar J$ and $C_X$, since the inherent symmetries of these functions are much simpler than those imposed on $\mandel{}$ by the kinematics.
$C$ ($C_3$) is (anti)symmetric under $\Z_{\{2,2,2\}}$ acting on its arguments, while $C_{11}$ and $C_{21}$ are symmetric or antisymmetric under various subgroups thereof.
The symmetries of $C_X$ and those of the stripped amplitudes interplay nontrivially, giving rise to several orbits.
Denoting by $i\cdots j(p)$ the orbit that has $p$ distinct elements including $C_X(p_i,\ldots,p_j)$, the orbits of $C$ and $C_3$ under various $\Z_R$ are
\begin{equation}
    \begin{aligned}
        \Z_6:\;& \left\{\scalemath{.7}{
            \begin{gathered}
                \mathbf{123456(2)},
                \mathbf{142536(1)},
                152436(3),
                \mathbf{162435(6)},\\
                \mathbf{162534(3)},
            \end{gathered}}\right.\\
        \Z_{\{2,4\}}:\;& \left.\scalemath{.7}{
            \begin{gathered} 
                \mathbf{123456(2)},
                123546(1),
                162435(4),
                \mathbf{162345(8)},
            \end{gathered}}\right.\\
        \Z_{\{3,3\}}:\;&\left.\scalemath{.7}{
            \begin{gathered} 
                \mathbf{123456(9)},
                \mathbf{142536(3)},
                \mathbf{162534(3)},
            \end{gathered}}\right. \\
        \Z_{\{2,2,2\}}:\;& \left.\scalemath{.7}{
            \begin{gathered} 
                \mathbf{123456(1)},
                162435(8),
                162534(6),
            \end{gathered}}\right.
    \end{aligned}
\end{equation}
and those of $C_{11}$ and $C_{21}$ are
\begin{equation}
    \begin{aligned}
        \Z_6:\;& \left\{\scalemath{.7}{
            \begin{gathered}
                \mathbf{123456(6)},
                \mathbf{235614(3)},
                253614(3),
                254613(6),\\
                263415(6),
                263514(3),
                264513(6),
                354612(6),\\
                364512(6),
            \end{gathered}}\right.\\
        \Z_{\{2,4\}}:\;& \left\{\scalemath{.7}{
            \begin{gathered} 
                123456(4),
                124635(2),
                162435(4),
                \mathbf{132456(8)},\\
                263415(8),
                263514(8),
                264513(8),
                \mathbf{354612(1)},\\
                \mathbf{345612(2)},
            \end{gathered}}\right.\\
        \Z_{\{3,3\}}:\;&\left.\scalemath{.7}{
            \begin{gathered} 
                123456(18),
                \mathbf{234516(9)},
                263415(9),
                263514(9),
            \end{gathered}}\right. \\
        \Z_{\{2,2,2\}}:\;& \left.\scalemath{.7}{
            \begin{gathered} 
                123456(3),
                263415(12),
                264513(24),
                364512(6).
            \end{gathered}}\right.
    \end{aligned}
\end{equation}
Only those marked in bold actually appear in the amplitude.
This can be understood from the limited arrangements of legs around the diagram \cref{fig:6piNLO:C} that produce $\flav_R(\f_1,\ldots,\f_6)$ as a flavor structure, as is clarified by the technology of \cref{app:groups}.

\section{The six-meson amplitude: expressions}
\label{app:6pi-res}

In this appendix, we explicitly present the subamplitudes $\dgampl{i}_R$, whose relation to the full amplitude is given in \cref{sec:ampl}.
Note that we only present the nonpole part here; the pole part is implicitly expressed in \cref{eq:pole-nonpole}.
We use kinematic variables $s_i,t_i,u_i$ ($i=1,\ldots,9$) which are motivated and defined in \cref{app:deorbit}.
With these, the LO amplitude of \cref{eq:6meson-LO} can be deorbited to
\begin{subequations}
    \begin{align}
        F_\pi^4 \dsgampl{6}{\LO,1}     &= \frac{M_\pi^2 - 2s_1 + s_9}{16},\\
        F_\pi^4 \dsgampl{3,3}{\LO,1}   &= \frac{9u_9 - 8M_\pi^2}{288n},
    \end{align}
\end{subequations}
and the NLO part is%
\footnote{
    In many of the terms below, deorbiting has been carried out with $\Z_R$ rather than $\Z_R^\rev$: Although the fully expanded expressions are shorter in the latter case, they turned out to factorize more neatly in the former, making it better for presentation.
    We have inserted appropriate factors of 2 so that the full amplitude is obtained using \cref{eq:deorbit} with $\Z_R^\rev$ in all cases (recall that $\tilde A$ is not unique).}$^,$%
\footnote{
    Note that we freely mix the deorbiting-optimized bases $s_i,t_i,u_i$ with each other, and with $p_i\cdot p_j$ and $\mandel{ij\cdots}$, whenever doing so allows the expressions to be written more compactly.
    Unlike deorbiting, this rewriting is not systematic and we do not claim that the result is optimal.}$^,$%
\footnote{
    Computer-readable versions of these expressions, as well as the stripped and full amplitudes and the programs used to obtain them, are available from the authors upon request.}
%NOTE: this is mostly pasted from the automatically formatted amplitude expressions, with some
% final touches added manually.
\begin{widetext}
    \begin{subequations}
        \begin{align}
            F_\pi^6 \dsgampl{6}{\NLO, 1} &=
                \smallfrac{M_\pi^4}{4 n}  \pig\{  \bar J(p_1,p_2) - L - \kappa \pig\}  % M6p4R6.A1N.001.hf
                \notag\\
                &+ \smallfrac{M_\pi^4}{8 n}     % M6p4R6.A1N.002.hf
                \pig\{
                    2C_{11}(p_1,\ldots,p_6) + 2 C(p_1,\ldots,p_6) \big[s_{7}-M_\pi^2\big] + C(p_1,p_6,p_2,p_5,p_3,p_4)\big[s_{8} -2 s_{6} + s_{9}\big] 
                \pig\} \notag\\
                &-  L_0^\r    % M6p4R6.A1N.005.hf
                \pig\{
                    2 M_\pi^4 + 4 M_\pi^2 ( s_{7}-2 s_{1} ) + s_{1} ( s_{1} + 2 s_{4} + 3 s_{5} + 2 s_{6} - 3s_7) - s_{7} ( 3 s_{2} + 2 s_{3} - s_{7} - s_{9} )
                \pig\}\notag\\
                &- \smallfrac{1}{4}  L_3^\r    % M6p4R6.A1N.006.hf
                \pig\{
                    7 M_\pi^4 - 2 M_\pi^2 (7s_{1}-4s_7) +  s_{1} (2s_{1} + 2s_{4} + 2s_{5}  - 3s_7 - 3s_9) + s_{7}^2
                \pig\}\notag\\
                &+  M_\pi^2 L_5^\r    % M6p4R6.A1N.007.hf
                \pig\{
                    2 M_\pi^2 - 2 s_{1} + s_{7}
                \pig\}
                -2 M_\pi^4  L_8^\r    % M6p4R6.A1N.008.hf
                \\
            F_\pi^6 \dsgampl{6}{\NLO,\sgn} &=
                -\smallfrac{1}{48}  C_{3}(p_1,\ldots,p_6)    % M6p4R6.A2N.016.hf
                \notag\\
                &- \smallfrac{1}{96}      % M6p4R6.A2N.012.hf
                \pig\{
                    \overline C_{21}(p_1,\ldots,p_6) \big[ 4 M_\pi^2 - s_{1} - s_{3} + 4 s_{5} + 2 s_{9} \big]
                    - 3 \overline C_{21}(p_2,p_3,p_5,p_6,p_1,p_4) \big[2 s_{5} - s_{7} - s_{8}\big]
                \pig\}\notag\\
                &-\smallfrac{1}{64}      % M6p4R6.A2N.014.hf
                \pig\{
                     C_{11}(p_1,\ldots,p_6) \big[ 2 M_\pi^2 (s_{1} + s_{3}) - s_{1} s_{3} \big]
                     + 8 C_{11}(p_3,p_5,p_4,p_6,p_1,p_2)\big[( p_3 \cdot p_5) (p_4\cdot p_6)\big]
                \pig\}\notag\\
                &+ \smallfrac{1}{384}  C(p_1,\ldots,p_6)    % M6p4R6.A2N.020.hf
                \pig\{
                    24 M_\pi^4 s_{1}+ 12 M_\pi^2 s_{1}(s_1 - s_3 - s_9 - s_7) - 3 (s_{1}^2 s_{3} + s_1 s_3^2) + 2 s_{1} s_{3} (2 s_{5}  + 3 s_{9})
                \pig\}\notag\\
                &+ \smallfrac{1}{128}  C(p_1,p_6,p_2,p_5,p_3,p_4)    % M6p4R6.A2N.017.hf
                \pig\{
                    s_{3} s_{6} (2 s_{6} - s_{8} - s_{9})
                \pig\}\notag\\
                &+ \smallfrac{1}{16}  C(p_1,p_6,p_2,p_4,p_3,p_5)    % M6p4R6.A2N.018.hf
                \pig\{
                    (p_2 \cdot p_4) (p_3 \cdot p_5) % (u_1 + u_3 - u_4 - u_6 + u_7 + u_8 - u_9)
                    (4 M_\pi^2 - 2 s_{1} - s_{2} - s_{4} - 2 s_{5} -2 s_{6} + 2 s_{7} + s_{8} + s_{9})
                \pig\}\notag\\
                &+ \smallfrac{1}{192}  C(p_1,p_4,p_2,p_5,p_3,p_6)    % M6p4R6.A2N.019.hf
                \pig\{
                    6 s_{3} s_{4} s_{5} + 2 s_{2} s_{4} s_{6} + 6 s_{1} ( s_{8}^2-s_{6} s_{7} ) 
                        \peqbreak{-} s_9\big[6 s_{2} s_{4} + 6s_{5} (s_{3} + s_{4})  + s_{8}\big( 2s_{7} -6 (s_{3} + s_{4} + s_{5}) + 3 (s_8+s_9)\big)\big]
                \pig\}\notag\\
                &- \smallfrac{1}{384}  \bar J(p_1,p_2)    % M6p4R6.A2N.023.hf
                \pig\{
                    32 M_\pi^4 - M_\pi^2 \big[ 2 s_{1} + 8 s_{2} + 5 s_{3} + 5 s_{5} + 2 (4 s_{6} + s_{7} - 2 s_{8} + s_{9})\big] 
                    \peqbreak{-} s_{1} (3 s_{1} - 2 s_{2} - 5 s_{3} - 12 s_{4} - 5 s_{5} - 2 s_{6} + 4 s_{7} + s_{8} +4 s_{9})
                \pig\}\notag\\
                &-\smallfrac{1}{32}  \bar J(p_1,p_3)    % M6p4R6.A2N.021.hf
                \pig\{
                    % (M_\pi^2 - s_1 - s_2 + s_7) (4 M_\pi^2 - 2 s_4 - 2 s_5 - s_6 - s_1 - s_2 - s_3 + 2 s_7 + s_8 + s_9)
                    (p_1\cdot p_3) (\mandel{46} + \mandel{135} - 5M_\pi^2)
                \pig\}\notag\\
                &+ \smallfrac{1}{256}  \bar J(p_1,p_4)    % M6p4R6.A2N.022.hf
                \pig\{
                    2 M_\pi^2 (2 s_{2} - s_7 - s_8) + s_{8} (2 s_{5} + 4 s_{1} + 4 s_{3} - s_{8} - 4 s_{9})
                        \peqbreak{-} 4 s_{2} (s_{1} + s_{3} + s_{4} + s_{6} - 2 s_{9}) + s_{7} (4 s_{1} + 2 s_{2} + 4 s_{3}  - s_7 - 2 s_{8} - 4 s_{9}) 
                \pig\}\notag\\
                &+ \smallfrac{L+\kappa}{384}     % M6p4R6.A2N.024.hf
                \pig\{
                    20 M_\pi^4 + M_\pi^2 ( 26 s_{7}-68 s_{1} ) + s_{1} (32 s_{4} + 54 s_{5} + 40 s_{6} + 9 s_{1} - 47 s_7)  
                        \peqbreak{+} s_{9} ( 27 s_{8} -47 s_{4} - 49 s_{5} + 15 s_{9})
                \pig\}\notag\notag\\
                &- \smallfrac{\kappa}{1152}    % M6p4R6.A2N.noprefix.hf
                \pig\{
                    10 M_\pi^4 - 2M_\pi^2 ( 23 s_{1} - 8 s_{9}) + s_{1} (16 s_{4} +18 s_{5}  +8s_{6} + 3 s_{1} - 16 s_7) 
                        \peqbreak{-} 2s_{7} (  8 s_{2} + s_{3} - 3 s_{7}) 
                \pig\}
                \\[3cm]
            F_\pi^6 \dsgampl{6}{\NLO,\kk} &=
                -\smallfrac{n}{48}  C_{3}(p_1,\ldots,p_6)    % M6p4R6.A3N.002.hf
                + \smallfrac{n}{96}  \overline C_{21}(p_1,\ldots,p_6)    % M6p4R6.A3N.000.hf
                \pig\{
                    2 M_\pi^2 + s_{1} + s_{3} - 4 s_{5} - 2 s_{9}
                \pig\}\notag\\
                &-\smallfrac{n s_{1} s_{3}}{64}  C_{11}(p_1,\ldots,p_6)    % M6p4R6.A3N.001.hf
                + \smallfrac{n}{384}  C(p_1,\ldots,p_6)    % M6p4R6.A3N.003.hf
                \pig\{
                    3 s_{1}^2 s_{3} + 3 s_{3} s_{5} (2 M_\pi^2 + s_{5}) - 2 s_{1} s_{5} (2 s_{3} + 3 s_{7})
                \pig\}\notag\\
                &- \smallfrac{n}{384}  \bar J(p_1,p_2)    % M6p4R6.A3N.004.hf
                \pig\{
                    20 M_\pi^4 - M_\pi^2 \big[ 5 s_1 + 8 s_{2} + 2 s_{3} + 2 s_{5} + 8 s_{6} - 4 (s_{7} + s_{8} + s_{9})\big] 
                        \peqbreak{+} s_{1} ( 9 s_{1} + 2 s_{2} - s_{3} - s_{5} + 2 s_{6} - 4 s_{7} - s_{8} - 4 s_{9}) 
                \pig\}\notag\\
                &+ \smallfrac{n(3L+\kappa)}{1152}      % M6p4R6.A3N.005.hf
                \pig\{
                    2 M_\pi^4 + M_\pi^2 ( 8 s_{9}-23 s_{1} )+ s_{1} (9 s_{1} + 20 s_{4} + 18 s_{5} + 4 s_{6} - 20 s_7) - s_{7} (20 s_{5} + s_{6} - 6s_7)
                \pig\}\notag\\
                &- \smallfrac{ n \kappa  }{384}   % M6p4R6.A3N.006.hf
                \pig\{
                    2 M_\pi^4 -  s_{1} (5 s_{1} + 8 s_{4} + 6 s_{5} - 8 s_7 - 8 s_9) - 2 s_{7}^2
                \pig\}
                \\
            F_\pi^6 \dsgampl{2,4}{\NLO, 1} &=
                \smallfrac{M_\pi^4}{8 n^2}  \pig\{ L+\kappa - 2 M_\pi^2 C(p_1,\ldots,p_6) - \bar J(p_1,p_2)\pig\}    % M6p4R24.A1N.000.hf
                \notag\\
                &+  L_1^\r    % M6p4R24.A1N.002.hf
                \pig\{
                     4 M_\pi^4 + t_{1} t_{6} + t_{4} t_{8} - 2 M_\pi^2 (t_{1} + 2 t_{8})
                \pig\}
                +  \smallfrac{1}{2} L_4^\r    % M6p4R24.A1N.004.hf
                \pig\{
                    2 M_\pi^2 (M_\pi^2 - t_{1} + t_{8})
                \pig\}\notag\\
                &+ \smallfrac{1}{16}  L_2^\r    % M6p4R24.A1N.003.hf
                \pig\{
                     16 M_\pi^4 - 8 M_\pi^2 (2 t_{1} + t_{5}) + 4 t_{1} t_{6} + 4 t_{4} t_{8} - t_{9}^2
                \pig\}
                -2 M_\pi^4  L_6^\r    % M6p4R24.A1N.005.hf
                \\
            F_\pi^6 \dsgampl{2,4}{\NLO,\kk} &= 
                 \smallfrac{1}{32}  \overline C_{21}(p_3,\ldots,p_6,p_1,p_2)    % M6p4R24.A3N.000.hf
                \pig\{
                    2 M_\pi^2 - t_{1} - t_{4}
                \pig\}
                -\smallfrac{1}{4}  C_{11}(p_1,p_3,p_2,p_4,p_5,p_6)    % M6p4R24.A3N.001.hf
                \pig\{
                    % (2 M_\pi^2 - t_{1} + t_{2} - t_{3} - t_{4} + t_{5} + t_{6} - t_{7} - t_{8} + t_{9}) (2 M_\pi^2 + t_{1} - t_{2} - t_{3} - t_{4} - t_{5} + t_{6} + t_{7} - t_{8} + t_{9}) 
                    (p_1\cdot p_3) (p_2\cdot p_4)
                \pig\}\notag\\
                &+ \smallfrac{1}{512}  C(p_1,\ldots,p_6)    % M6p4R24.A3N.003.hf
                \pig\{
                    32 M_\pi^6 - 32 M_\pi^4 (t_{1} + t_{2} + t_{5}) 
                    \peqbreak{-} t_{1}^2 (t_{3} + t_{4} + 2 t_{5}) - t_{2} \big[t_{2} t_{4} + 2 (t_{2} + t_{4}) t_{5}\big] - 2 t_{4} t_{5} t_{7} - 2 t_{4} t_{7}^2 
                    \peqbreak{+} 2 M_\pi^2 \big[2 t_{1}^2 + 4 t_{1} t_{2} + 2 t_{2}^2 + 3 t_{2} (t_{4} + 4 t_{5}) +3 t_{1} (t_{3} + 2 t_{4} + 4 t_{5}) + 2 t_{7} (t_{5} + t_{7})\big] 
                    \peqbreak{-}t_{1} \big[t_{4}^2 + 4 t_{4} t_{5} + 2 t_{2} (t_{3} + t_{4} + 2 t_{5}) + 2 t_{5} (t_{3} + t_{5} + t_{7})\big]
                \pig\}\notag\\
                &+ \smallfrac{1}{8}  C(p_1,p_6,p_2,p_3,p_4,p_5)    % M6p4R24.A3N.002.hf
                \pig\{
                    (p_1\cdot p_6) (p_2\cdot p_3) (2 M_\pi^2 + s_2 - 2 s_8 - 2 s_4 + s_6)
                    % (2 M_\pi^2 - t_1 - t_2 - t_3 + t_4 + t_5 + t_6 - t_7 - t_8 - t_9) (2 M_\pi^2 - t_1 - t_2 + t_3 - t_4 - t_5 + t_6 + t_7 - t_8 - t_9)\big[ 22 M_\pi^2 - t_1 - t_2 - 2 t_3 - 2 t_4 - 2 t_5 - 3 t_6 - 2 t_7 - t_8 - t_9\big]
                  %  (2 s_2 - s_3 - 2 s_4 - s_8 + s_9)(p_1 \cdot p_6) (p_2 \cdot p_5)
                \pig\}\notag\\
                &+ \smallfrac{1}{128}  \bar J(p_1,p_2)    % M6p4R24.A3N.005.hf
                \pig\{
                     M_\pi^2 (6 t_{1} - 2 t_{5}) + t_{1} ( t_{6}-2 t_{2} - t_{3} ) + t_{4} ( t_{8}-t_{4} ) -2 M_\pi^4 
                \pig\}\notag\\
                &-\smallfrac{1}{16}  \bar J(p_1,p_3)    % M6p4R24.A3N.004.hf
                \pig\{
                    (p_1\cdot p_3) (\mandel{46} + \mandel{246} - 5 M_\pi^2)
                    %(22 M_\pi^2 - 2 t_2 - t_3 - 2 t_4 - t_1 - 2 t_8 - 2 t_5 - 2 t_6 - 2 t_7 + t_9)
                    % (2 M_\pi^2 - t_2 - t_3 - t_4 + t_1 - t_8 - t_5 + t_6 + t_7 + t_9) 
                    % (p_1\cdot p_3) \big[\mandel{46} + \mandel{246} - 6 M_\pi^2\big]
                \pig\}\notag\\
                &+ \smallfrac{L+\kappa}{256}     % M6p4R24.A3N.006.hf
                \pig\{
                    4 M_\pi^2 (t_{1} + 3 t_{5}) + 2 t_{1} (t_{1} + 2 t_{2} + t_{3} - 3 t_{6}) - 6 t_{4} t_{8} + t_{9}^2 - 12 M_\pi^4
                \pig\}
                \\
            F_\pi^6 \dsgampl{3,3}{\NLO, 1} &= 
                \smallfrac{M_\pi^4}{6 n^2}  \pig\{ L+\kappa - \bar J(p_1,p_2)\pig\}    % M6p4R33.A1N.000.hf
                - \smallfrac{1}{16 n}  ( 4 L_0^\r + L_3^\r )     % M6p4R33.A1N.001.hf
                \pig\{
                    4 M_\pi^2 u_1  + 2u_1(u_4 + u_5) - u_9(4u_8 + 3u_9)
                    % 4 M_\pi^4 + 8 M_\pi^2 u_{1} - 2 u_{1} (u_{4} + u_{5}) + u_{9} (4 u_{8} + 3 u_{9})
                \pig\}\notag\\
                &- \smallfrac{1}{2n}  L_3^\r    % M6p4R33.A1N.002.hf
                \pig\{
                    u_{1} (u_{4} + u_{5}) - u_{9} (u_{8} + u_{9})
                \pig\}
                + \smallfrac{ 1}{6 n}  L_5^\r  % M6p4R33.A1N.003.hf
                \pig\{
                    M_\pi^2 ( 3 u_{9}- 8 M_\pi^2 )
                \pig\}
                +\smallfrac{4 M_\pi^4}{3}  \pig\{\smallfrac{1}{n} L_8^r - \smallfrac{1}{3} L_7^\r\pig\}    % M6p4R33.A1N.004.hf
                % + \smallfrac{4 M_\pi^4}{3 n}  L_8^\r    % M6p4R33.A1N.005.hf
                \\
            F_\pi^6 \dsgampl{3,3}{\NLO,\sgn} &= 
                 \smallfrac{M_\pi^4}{24 n}  \bar J(p_1,p_2)    % M6p4R33.A2N.000.hf
                -\smallfrac{6L+5\kappa}{2304n}    % M6p4R33.A2N.001.hf
                \pig\{
                    16 M_\pi^4 - 12 M_\pi^2 u_{1} + 2 u_{1} (u_{4} + u_{5}) + u_{9} (4 u_{8} + u_{9})
                \pig\}
                -\smallfrac{M_\pi^4 \kappa}{144 n}      % M6p4R33.A2N.002.hf
                \\
            F_\pi^6 \dsgampl{3,3}{\NLO,\sgnp} &= 
                \smallfrac{u_{8}-2M_\pi^2}{32}  \overline C_{21}(p_2,\ldots,p_5,p_1,p_6)    % M6p4R33.A4N.000.hf
                \notag\\
                &+ \smallfrac{1}{384}  C(p_1,p_6,p_2,p_5,p_3,p_4)    % M6p4R33.A4N.001.hf
                \pig\{
                    8M_\pi^6 - 12M_\pi^4 u_8 + 6 M_\pi^2 u_4 u_8 - u_3 u_4 u_8
                \pig\}\notag\\
                &- \smallfrac{1}{384} C(p_1,p_4,p_2,p_5,p_3,p_6)
                \pig\{
                     8 M_\pi^6 - 12 M_\pi^4 u_7 + 6 M_\pi^2 u_4 u_7 - u_1 u_4 u_7    % M6p4R33.A4N.002.hf
                \pig\}\notag\\
                &- \smallfrac{1}{128}  \bar J(p_1,p_4)    % M6p4R33.A4N.003.hf
                \pig\{
                    (p_1\cdot p_4) (u_{4} + u_{7} - 2 u_{9})
                \pig\}
                - \smallfrac{ L+\kappa}{128}    % M6p4R33.A4N.004.hf
                \pig\{
                    u_{1} u_{5} - u_{6} u_{9}
                \pig\}
                \\
            F_\pi^6 \dsgampl{3,3}{\NLO,\kk} &= 
                -\smallfrac{p_1\cdot p_6}{8}  \overline C_{21}(p_2,\ldots,p_5,p_1,p_6)\notag\\
                &+ \smallfrac{1}{256}  C(p_1,\ldots,p_6)    % M6p4R33.A3N.002.hf
                \pig\{
                    2(p_3\cdot p_4) \big[ 64 M_\pi^4 - 2 (u_{1} + u_{2})^2 + u_{3}^2 + 2 u_{4}^2 + 2 u_{1} u_{5} + 4 u_{4} u_{6} + u_{6}^2 \pppeqbreak{-}2 u_3 (2 u_4 + u_6) + 2 u_7 (2 u_{1} + u_{2}) 
                    + 2 u_{4} u_{8} + 16 M_\pi^2 (u_{3} - 2 u_{4} - u_{6})\big]  
                    \ppeqbreak{+} u_9\big[32M_\pi^4-4M_\pi^2(3u_3+2u_4+u_6) \pppeqbreak{+} 2u_3(u_6-u_3+2u_8) + u_9(u_3-2M_\pi^2)\big]
                \pig\}\notag\\
                &- \smallfrac{1}{96}  C(p_1,p_6,p_2,p_5,p_3,p_4)    % M6p4R33.A4N.001.hf
                \pig\{
                    8M_\pi^6 - 12M_\pi^4 u_8 + 6 M_\pi^2 u_4 u_8 - u_3 u_4 u_8
                \pig\}\notag\\
                &+ \smallfrac{1}{384}  \bar J(p_1,p_2)    % M6p4R33.A3N.003.hf
                \pig\{
                    32 M_\pi^4 - 36M_\pi^2(u_1+u_2-u_3)
                    + 3 \big[ u_{1}^2 + u_{3}^2 + 2 u_{3} u_{5} - 2 u_{2} (u_{3} - u_{4} + u_{5} + u_{7}) \pppeqbreak{+} 2 u_{8} u_{9} +u_{9}^2
                    + 2 u_1 (u_4 - u_3 - u_5 + u_6 - u_7 + u_9)\big]
                \pig\}\notag\\
                &+ \smallfrac{1}{64}  \bar J(p_1,p_4)    % M6p4R33.A3N.003.hf
                \pig\{
                    (p_1\cdot p_4)(u_4 + u_7 - 2 u_9)
                \pig\}\notag\\
                &+ \smallfrac{3L+2\kappa}{2304}     % M6p4R33.A3N.004.hf
                \pig\{
                    32M_\pi^4 - 12 M_\pi^2 u_{1} +2 u_{1} (u_{4} + 13u_{5}) - u_{9} (20 u_{8} + 17 u_{9})
                \pig\}\notag\\
                &+ \smallfrac{\kappa}{576}    % M6p4R33.A3N.005.hf
                \pig\{
                    8M_\pi^4 - 6M_\pi^2 u_1 + u_1(u_4 + 7 u_5) - 4u_9(u_8+u_9)
                \pig\}
                \\
            F_\pi^6 \dsgampl{2,2,2}{\NLO, 1} &= 
                 \smallfrac{M_\pi^6}{6 n^3}  C(p_1,\ldots,p_6)    % M6p4R222.A1N.000.hf
        \end{align}
    \end{subequations}
    \pagebreak
\end{widetext}
\noindent Here, $\bar J(p_i,p_j) \equiv \bar J\big((p_i+p_j)^2\big)$.
A few features can be observed:
\begin{itemize}
    \item The NLO LECs only appear in $\gampl{\NLO,1}$.
    \item $\gampl{\NLO,\sgnp}_R$ only exists for $R=\{3,3\}$.
    \item $\gampl{\NLO,\kk}_R$ and $\gampl{\NLO,\sgnp}_R$ are independent of $n$ for $R\neq\{6\}$, while $\sgampl{6}{\NLO,\kk}$ is proportional to $n$ and is the only place where positive powers of $n$ appear.
    \item $\sgampl{2,2,2}{\NLO,1}$ is the only $R=\{2,2,2\}$ subamplitude, is proportional to $n^{-3}$, and is the only place where this power appears.
\end{itemize}
These features and their generalizations are derived in \cref{app:groups}.

\section{Symmetries and group-dependent features of the amplitude}
\label{app:groups}

In this appendix, we derive the features described in the previous section using the technique we here dub \emph{diagrammatic flavor-ordering}, wherein modified Feynman diagrams allow direct calculation of stripped amplitudes without going through the full amplitude.
Simpler cases of the technique have been used for a long time~\cite{Osborn:1969ku,Susskind:1970gf,Kampf:2013vha}, but the extension beyond LO and $R=\{k\}$ is more recent~\cite{masterthesis,Bijnens:2019eze}.
A somewhat similar approach can be found in \rcite{Low:2019ynd}.
In the preparation of this paper, we refined the technique and performed the first loop calculations using it, but it turned out that the proliferation of diagrams caused by the inclusion of loops and nonzero masses --- nearly 200 distinct topologies compared to 9 without flavor-ordering --- outweighed any efficiency advantages the technique had over standard Feynman diagrams, rendering it impractical for our purposes.
Nevertheless, the manifest relation between kinematics and flavor structure in flavor-ordered diagrams can be used to illuminate some features that are obscured with the standard approach.

\subsection{Diagrammatic flavor-ordering}
Here, we give a brief summary of this technique; see \rcite{Bijnens:2019eze} for a detailed version, and \rcite{masterthesis} for one including loops.

By `flavor-ordered', we mean a quantity whose flavor structure is $\flav_R$ for some $R$, i.e.\ whose flavor indices are in natural order (up to $\Z_R$).
Such a quantity is invariant under $\Z_R$ acting simultaneously on its flavor indices and momenta.
The stripped amplitude is obtained by keeping only the flavor-ordered parts of the amplitude and then dropping the flavor structure.
Diagrammatic flavor-ordering is based on the observation that the Fierz identity, \cref{eq:fierz-SU,eq:fierz-SOp}, generally preserves flavor-ordering: if two sub-diagrams contain $\tr{t^aA}$ and $\tr{t^aB}$, joining them will result in $\tr{AB}, \tr{A}\tr{B}$ or $\tr{AB^\R}$, all of which keep the (possibly reversed) order of flavor indices in $A$ and $B$.
Therefore, a diagram is flavor-ordered only if its sub-diagrams, all the way down to the vertices, are flavor-ordered.

\begin{figure}[t]
    \centering
    \begin{subfigure}[t]{0.32\columnwidth}
        \includegraphics{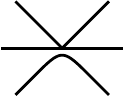}
        \caption{$1\times$}
        \label{fig:6piNLO-split:24point}
    \end{subfigure}
    \begin{subfigure}[t]{0.32\columnwidth}
        \includegraphics{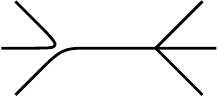}
        \caption{$4\times$}
        \label{fig:6piNLO-split:24prop}
    \end{subfigure}
    \begin{subfigure}[t]{0.32\columnwidth}
        \includegraphics{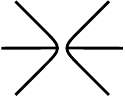}
        \caption{$1\times$}
        \label{fig:6piNLO-split:33}
    \end{subfigure}
    \caption{
        Flavor-ordered variants of \cref{fig:6piNLO:point,fig:6piNLO:prop} with split vertices.
        Note that the multiplicities given here refer to permutations in $\Z_R$.
        Diagrams (a) and (b) have flavor structure $\flav_{\{2,4\}}$ and contain $L_i$, $i=1,2,4,6$.
        Diagram (c) has $\flav_{\{3,3\}}$ and contains $L_7$.
        Flavor-ordered diagrams identical to \cref{fig:6piNLO:point,fig:6piNLO:prop,fig:6piNLO:2prop} have $\flav_{\{6\}}$, contain $L_i$, $i=0,3,5,8$, and have multiplicity $1\times$, $6\times$ and $3\times$, respectively.
        }
    \label{fig:6piNLO-split}
\end{figure}

We now desire a set of modified diagram-drawing rules that make diagrams inherently flavor-ordered, with the flavor structures manifest from the graphical shape of the diagram.
We think of each external leg as labeled by an index $i$, corresponding to momentum $p_i$ and flavor index $\f_i$.
We will call two legs \emph{flavor-connected} if their flavor indices reside in the same trace in the flavor structure.

For single-vertex diagrams, we indicate the flavor structure by adding gaps in the vertex between the groups of legs that are not flavor-connected, as is done in \cref{fig:6piNLO-split:24point,fig:6piNLO-split:33}.
We treat the vertices in multi-vertex diagrams like \cref{fig:6piNLO-split:24prop} similarly.
Since the two terms on the right-hand side of \cref{eq:fierz-SU:tree} treat flavor structures differently, we represent them by different propagators as if there were two species of particles: ordinary (solid line) and singlets (dashed line), the latter of which carry a factor of $-\frac1{\kk n}$.%
\footnote{
    The name ``singlet'' stems from how adding a singlet field $\phi^0$, whose associated generator $t^0=\frac{\1}{\sqrt{\kk n}}$ commutes with all $t^a$, results in the removal of the $1/n$ terms from the Fierz identity since e.g.\ $\tr{t^0A}\tr{t^0B} = \frac{1}{\kk n}\tr{A}\tr{B}$.
    Thus, the $1/n$ terms can be interpreted as the subtraction of diagrams with internal singlet lines, allowing other contractions to be done using only the $n$-independent terms.
    The singlet decouples from the other fields in $\tr{u_\mu u^\mu}$, so LO singlet vertices stem from $\tr{\chi_+}$ and therefore depend on the mass but not on the momenta [this is easiest to see in the exponential parametrization~\eqref{eq:u_exp}].
    This simplifies LO and NLO singlet diagrams and causes them to vanish in the massless limit.}

\begin{figure}[t]
    \centering
    \begin{subfigure}[t]{0.49\columnwidth}
        \includegraphics{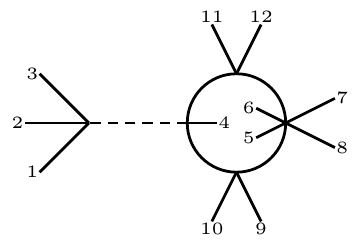}
        \caption{}
        \label{fig:read:indices}
    \end{subfigure}
    \begin{subfigure}[t]{0.49\columnwidth}
        \includegraphics{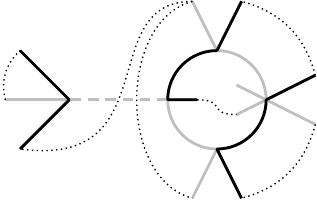}
        \caption{}
        \label{fig:read:paths}
    \end{subfigure}
    
    \begin{subfigure}[t]{0.49\columnwidth}
        \includegraphics{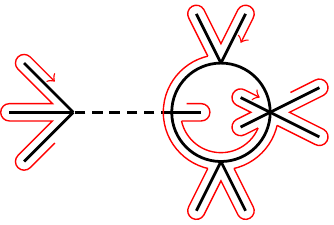}
        \caption{}
        \label{fig:read:zeta}
    \end{subfigure}
    \begin{subfigure}[t]{0.49\columnwidth}
        \includegraphics{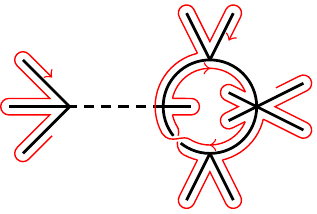}
        \caption{}
        \label{fig:read:xi}
    \end{subfigure}
    \caption{An illustration of how indices are read from a flavor-ordered diagram.
        (a) The diagram with indices assigned (up to $\Z_{\{3,3,6\}}$).
        (b) Examples of paths connecting legs without intersecting the diagram (dotted) and paths within the diagram (solid, with the rest of the diagram grayed out).
        (c) Paths showing the order in which the indices are read to give the indexing in (a).
        (d) The alternative path giving the $\sgn/2$ term under $\SOp$.
        Note that this results in a different indexing than in (a) and flavor structure $\flav_{\{3,9\}}.$}
    \label{fig:read}
\end{figure}

When combined with \cref{eq:fierz-SU:loop}, these rules allow the flavor structure of any $\SU$ diagram to be read off, as illustrated in \cref{fig:read}.
Two legs are flavor-connected if and only if the following conditions hold:
\begin{itemize}
    \item They are joined by an uninterrupted path through the diagram (vertex gaps and singlet propagators interrupt it).
    \item They can be joined by a line that does not intersect the diagram at any point (it can pass through vertex gaps and singlet propagators).
\end{itemize}
This is illustrated in \cref{fig:read:paths}.
    
To read the indices, follow the outline of each flavor-connected set of legs, keeping the diagram to the right of the path (thus reading the indices of tree diagrams in clockwise order), as illustrated in \cref{fig:read:zeta}.%
\footnote{
    These rules become more complicated at two-loop level and above, where non-planar diagrams may appear.
    However, all diagrams can be drawn without self-intersections on a surface of sufficiently high topological genus (planar diagrams on a sphere, non-planar two-loop diagrams on a torus, etc.).
    One must then imagine the diagram drawn on such a surface (but not one of higher genus than necessary) when determining flavor-connectedness or assigning indices.}
The starting point is arbitrary due to $\Z_R$ symmetry.
When a loop is `empty', like those in \cref{fig:6piNLO-A233}, a factor of \mbox{$\tr{\1}=\kk n$} is added.

For the purposes of momentum flow, flavor-ordered diagrams are treated just like ordinary diagrams, and the two kinds of propagators are kinematically identical.
However, flavor-ordered diagrams are typically sensitive to the order in which legs are arranged around a vertex (see e.g.\ \cref{fig:6piNLO-A433}).
This, along with the combination of singlet and ordinary propagators, leads to the proliferation mentioned earlier.
All diagrams must be summed over $\Z_R$ (with appropriate symmetry factors) and added up to obtain $A_R$.

The above rules hold for $\SU$, and to a large extent also for $\SOp$.
In fact, the cases where they are equivalent (up to the substitution $n\to\kk n$) exactly correspond to $\gampl{1}$ of \cref{eq:group-univ}.

\subsection{Differences between the groups}\label{sec:diff}
We will now discuss all contexts in which differences between $\SU$ and $\SOp$ may arise.
The following fully accounts for the patterns seen in the six-meson amplitude:

\paragraph{Tree diagrams.}\label{par:tree} View a $\SOp$ diagram as being built by adding vertices one by one. 
With $A$ belonging to the partially completed diagram and $B$ to the vertex, $\tr{t^aA}\tr{t^aB}\to\tfrac12[\tr{AB} + \tr{AB^\R}]$ gives one flavor-ordered term and one that is discarded.
Adding a structurally identical diagram but with some indices permuted so that $B$ is reversed gives $\tr{t^aA}\tr{t^aB^\R}\to\tfrac12[\tr{AB^\R} + \tr{AB}]$: Again, one term is kept and one discarded.
However, $\tr{t^aB} = \tr{t^a B^\R}$ under $\SOp$, so the kinematic structure of the vertex must be invariant under that index permutation. 
Thus, the two flavor-ordered terms are identical and add up to the same $\tr{AB}$ given by $\SU$.
This proves, to all orders in the chiral counting, that \cref{eq:fierz-SU:tree,eq:fierz-SOp:tree} fail to introduce any differences between $\SU$ and $\SOp$.
In other words, \emph{$\SU$ and $\SOp$ are equivalent at tree level (up to $n\to\kk n$)}, so all tree diagrams go into $\gampl{1}$.
The only caveat is if any differences are introduced at the Lagrangian level, but this happens first at NNLO (see below).

\begin{figure}[t]
    \centering
    \begin{subfigure}[t]{0.32\columnwidth}
        \includegraphics{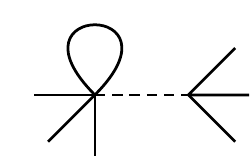}
        \caption{$18\times$}
        \label{fig:6piNLO-A233:A-prop}
    \end{subfigure}
    \begin{subfigure}[t]{0.32\columnwidth}
        \includegraphics{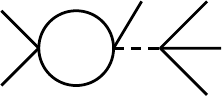}
        \caption{$18\times$}
        \label{fig:6piNLO-A233:B}
    \end{subfigure}
    \begin{subfigure}[t]{0.32\columnwidth}
        \includegraphics{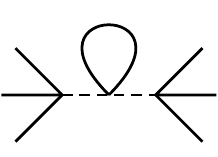}
        \caption{$9\times$}
        \label{fig:6piNLO-A233:A-2prop}
    \end{subfigure}
    \caption{
        A few of the flavor-ordered diagrams that contribute to $\sgampl{3,3}{\sgn}$.
        The multiplicities refer to permutations in $\Z_{\{3,3\}}$.
        All diagrams contain a single factor $1/n$ from singlets; in (c) the $1/n$ from the second singlet propagator is canceled by $\tr{AB^\R} = \tr{\1}=n$ (such a factor appears whenever a loop is not flavor-connected to any external leg).
        In their contributions to $\gampl{\kk}$, all $n$-dependence is canceled by $\tr{A}\tr{B} = \tr{A}\tr{\1}$ for (a,b) and $\tr{A}\tr{B} = \tr{\1}\tr{\1}$ for (c).}
    \label{fig:6piNLO-A233}
\end{figure}

\paragraph{Loops.}\label{par:loop} Viewing the loop as being formed by joining two legs of a tree diagram, we see that \cref{eq:fierz-SU:loop,eq:fierz-SOp:loop} must be applied if those legs are part of the same trace.
The term $\tr{t^aAt^aB}\to \tr{A}\tr{B}$ is the same in $\SU$ and $\SOp$, up to a factor $\frac1\kk$, giving rise to $\gampl{\kk}$.
The term $\tr{t^aAt^aB}\to \tr{AB^\R}$ is unique to $\SOp$ and contains $\pm$, giving rise to $\gampl{\sgn}$.%
\footnote{
    This necessitates the extra rule illustrated in \cref{fig:read:xi}.
    In detail, it is as follows: Starting from the regular index-assigning path [\cref{fig:read:zeta}], create arbitrarily located `gaps' in loop propagators until all legs can be joined by lines that do not intersect the diagram.
    Following the path, split the diagram into two terms every time a gap is reached.
    In one term, continue on the original path and multiply by $1/\kk$.
    In the other, go through the gap and multiply by $\sgn/2$.
    Do this recursively if you come across another gap, but the second time a given gap is reached, go through it unconditionally.
    Every time you go through a gap, swap the orientation of your path (clockwise to counterclockwise, and vice versa).}
Since this term gives a single trace, it almost always results in $R=\{6\}$, which explains why $\sgampl{6}{\sgn}$ is the longest subamplitude.
A few singlet diagrams, including those in \cref{fig:6piNLO-A233}, give $\sgampl{3,3}{\sgn}$ instead.

When $B=\mathbbb{1}$ in the $\gampl{\kk}$ case, we get a factor of $n=\tr{\1}$; this corresponds graphically to an `empty' loop as mentioned above.
This is the only source of positive powers of $n$, and explains why they only appear in $\gampl{\kk}$.
Those diagrams still contribute to $\gampl{\sgn}$ without a factor of $n$.

\paragraph{Singlets.}\label{par:singlet} The $\frac1{\zeta n}$ terms of the Fierz identity are the same for $\SU$ and $\SOp$, so singlet propagators behave the same in both cases.
When a singlet is part of a loop, one can let the singlet propagator `close' the loop, thereby avoiding all differences stemming from \cref{eq:fierz-SU:loop,eq:fierz-SOp:loop}.
Therefore, such diagrams go into $\gampl{1}$.
This does not apply when a singlet is outside a loop, but for our amplitude this only happens with $R=\{3,3\}$ diagrams like in \cref{fig:6piNLO-A233} and variations thereof.
The `empty' loop cancels the $n$-dependence in their contributions to $\gampl{1}$ and $\gampl{\kk}$.
Therefore, negative powers of $n$, which only arise from singlets, only show up in $\gampl{1}$ and (due to these diagrams) $\sgampl{3,3}{\sgn}$.

\begin{figure}[t]
    \centering
    \begin{subfigure}[t]{0.32\columnwidth}
        \includegraphics{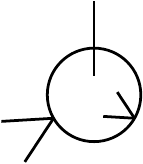}
        \caption{$9\times$}
        \label{6piNLO-A433:1}
    \end{subfigure}
    \begin{subfigure}[t]{0.32\columnwidth}
        \includegraphics{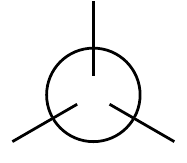}
        \caption{$9\times$}
        \label{6piNLO-A433:2}
    \end{subfigure}
    \begin{subfigure}[t]{0.32\columnwidth}
        \includegraphics{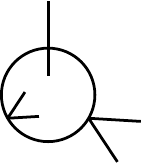}
        \caption{$9\times$}
        \label{6piNLO-A433:3}
    \end{subfigure}
    \caption{
        A few of the flavor-ordered diagrams that contribute to $\sgampl{3,3}{\sgnp}$.
        The multiplicities are for $\SU$; $\SOp$ permits more permutations, which is exactly why they give $\sgampl{3,3}{\sgnp}$.}
    \label{fig:6piNLO-A433}
\end{figure}

\paragraph{Trace-reversal.}\label{par:trace} The greatest differences come from the fact that $\tr{t^at^b\cdots t^c}=\tr{t^ct^b\cdots t^a}$ under $\SOp$ but not $\SU$.
{\em Therefore, $\ampl_R$ is invariant under reversal of individual traces under $\SOp$, but only under simultaneous reversal of all traces under $\SU$.}
Among the cases considered here, these types of reversal are only inequivalent when $R=\{3,3\}$.%
\footnote{
    In general, $R$ must contain at least two elements greater than 2, since $\tr{t^at^b}=\tr{t^bt^a}$ under all groups.}
For instance, in the diagrams in \cref{fig:6piNLO-A433}, the ``inside'' of the loop can be read both clockwise and counterclockwise under $\SOp$, but only one way under $\SU$, and the momentum dependence will be correspondingly different.
Such disorganized difference is what gives rise to $\sgampl{3,3}{\sgnp}$.
In all $R=\{3,3\}$ diagrams involving singlets, at least one trace can be reversed as a symmetry of the diagram (i.e.\ reversing a single-trace vertex), so they do not contribute to $\sgampl{3,3}{\sgnp}$.

\subsection{More particles, higher orders}
The patterns discussed here are straightforward to generalize.
At N$^\ell$LO (i.e.\ $\ell$ loops), \cref{eq:group-univ} becomes
\begin{equation}\label{eq:group-univ-gen}
    \ampl = \bigg\{\gampl{1} + \sgn\gampl{\sgn} + \sgnp\gampl{\sgnp} + \sum_{j=1}^\ell\frac{\gampl{\kk^j}}{\kk^j}\bigg\}_{\stackrel{}{n\to\kk n}}\,,
\end{equation}
since each loop can give another factor of $\frac1\kk$.
Most of the features discussed above remain, although some are softened: Negative powers of $n$ can appear in $\gampl{\kk^j}$ and $\gampl{\sgnp}$, and positive powers of $n$ in most subamplitudes except $\gampl{1}$.
$\gampl{\sgnp}_R$ exists for $R=\{2,3,3\}$, $\{4,4\}$, etc.
In an N$^\ell$LO amplitude, $\ampl_R$ with $|R|>\ell+1$ requires singlets breaking loops, which severely restricts the structure of that subamplitude; specifically, if $R=\{2,\ldots,2\}$, it will be similarly simple to our $\sgampl{2,2,2}{1}$.

As mentioned above, the equivalence of $\SU$ and $\SOp$ at tree level can only be broken by Lagrangian effects.
The first such effect is in the NNLO Lagrangian $\mathcal L^{(6)}$~\cite{Bijnens:1999sh}, where the 59th term $\tr{u_\mu u_\nu u_\rho}\tr{u^\mu u^\nu u^\rho}$ and the 61st term $\tr{u_\mu u_\nu u_\rho}\tr{u^\rho u^\nu u^\mu}$ are distinct under $\SU$ but equal under $\SOp$.
(The N$^3$LO Lagrangian $\mathcal L^{(8)}$~\cite{Bijnens:2018lez} contains several such cases.)
This only results in additional relations between the LECs; the functional form of $A$ is retained, and \cref{eq:group-univ-gen} remains valid, albeit a bit more redundant.

\renewcommand{\raggedright}{}
%\bibliography{references}

\providecommand{\href}[2]{#2}\begingroup\raggedright\endgroup

\end{document}